\documentclass{aa}  
\usepackage{graphicx}
\usepackage{txfonts}
\usepackage{natbib}
\bibpunct{(}{)}{;}{a}{}{,}
\begin{document}
   \title{The new Toulouse-Geneva Stellar Evolution Code including radiative accelerations of heavy elements}


   \author{S. Th\'eado \inst{1}\inst{2}
          \and
          G. Alecian \inst{3}
          \and
          F. LeBlanc \inst{4}
          \and
          S. Vauclair \inst{1}\inst{2}
          }

   \institute{Universit{\'e} de Toulouse; UPS-OMP; IRAP; Toulouse, France\\
               \email{sylvie.theado@ast.obs-mip.fr}
         \and 
             CNRS; IRAP; 14, avenue Edouard Belin, F$-$31400 Toulouse, France
         \and
             LUTH, Observatoire de Paris, CNRS, Universit{\'e} Paris Diderot,
             5 Place Jules Janssen, 92190 Meudon, France\\
               \email{georges.alecian@obspm.fr}
         \and
             D\'epartement de Physique et d'Astronomie, Universit\'e de Moncton, Moncton, NB, E1A 3E9, Canada\\
               \email{francis.leblanc@umoncton.ca}
             }

   \date{Received September 15, 2011; accepted}


  \abstract
   {Atomic diffusion has been recognized as an important process that has to be considered in any computations of stellar models. In solar-type and cooler stars, this process is dominated by gravitational settling, which is now included in most stellar evolution codes. In hotter stars, radiative accelerations compete with gravity and become the dominant ingredient in the diffusion flux for most heavy elements. Introducing radiative accelerations into the computations of stellar models modifies the internal element distribution and may have major consequences on the stellar structure. Coupling these processes with hydrodynamical stellar motions has important consequences that need to be investigated in detail.}
   {We aim to include the computations of radiative accelerations in a stellar evolution code (here the TGEC code) using a simplified method (SVP) so that it may be coupled with sophisticated macroscopic motions. We also compare the results with those of the Montreal code in specific cases for validation and study the consequences of these coupled processes on accurate models of A- and early-type stars.}
   {We implemented radiative accelerations computations into the Toulouse-Geneva stellar evolution code following the semi-analytical prescription proposed by Alecian and LeBlanc. This allows more rapid computations than the full description used in the Montreal code.}
   {We present results for A-type stellar models computed with this updated version of TGEC and compare them with similar published models obtained with the Montreal evolution code. We discuss the consequences for the coupling with macroscopic motions, including thermohaline convection.}
   {}


   \maketitle
%

\section{Introduction}

Accurate stellar modeling has recently been given a new boost with the advent of asteroseismology. The observations of oscillating stars and the analysis of the stellar oscillation properties brought new and powerful constraints on stellar models and allowed major progress in our understanding of stellar internal structure. Furthermore, since the discovery of the first extrasolar planets \citep{Wolszczan92,Mayor95}, the spectacular development of the exoplanet research field has also sparked renewed interest in stellar physics, the accurate knowledge of the host star being a necessary condition for characterizing the surrounding planets.

One of the most important successes of astrophysics was the understanding of the basics of stellar internal structure and evolution. This progress was supported by the computations of numerical models. Growing computational resources contributed to refining the description of the physics of the stellar medium (equation of state, opacities, nuclear reactions, etc) and allowed building a simplified but efficient and widely used ``standard model''. However, this standard model does not take the effects of rotation and magnetic fields or the occurrence of accretion or mass loss into account, and it considers convection as the only chemical transport process. 

Observations of chemical abundance anomalies in stars and unexpected stellar seismic behaviors have proved the necessity of including ``non standard processes'' in stellar evolutionary computations. In this framework, the main challenges encountered today by stellar physicists are to better determine the effects of rotation and magnetic fields and understand the chemical transport processes better. This last point, and more specifically modeling of atomic diffusion including the radiative accelerations on individual elements, represents a key ingredient for accurate stellar modeling.

The importance of atomic diffusion inside stars is well established: not only can it modify the atmospheric abundances \citep{Michaud70}, as observed in the so-called ``chemically peculiar stars'', but it can also have strong implications for the stellar internal structure \citep[e.g.][]{Richard01}. In main-sequence solar-type and cooler stars (below about 1.2 M$_{\odot}$), the radiative accelerations on the heavy elements are generally slower than gravity in absolute value, owing to the small radiation flux compared to hotter stars \citep{MichaudMiChVaetal1976}. The elements heavier than hydrogen sink, even if the efficiency of this sinking may be modulated by the radiative accelerations \citep{Turcotte98}. 

One of the biggest success of helioseismology was to prove the importance of atomic diffusion in the Sun. Its introduction in solar evolutionary models has significantly improved the agreement between the sound speed profile inside solar models and that deduced from helioseismic inversion techniques \citep[e.g.][]{JCD96,Richard96,Brun98,Turcotte98,Schlattl02}. (This agreement has however been spoiled by the new abundances proposed by Asplund et al. 2005, 2009.) Gravitational settling is now introduced in most stellar evolution codes. 

The effects of atomic diffusion on the seismic frequency of main-sequence, solar-type star models have been studied by \cite{Theado05}. They have shown that element segregation significantly  alters the internal structure of the models and their oscillations frequencies. The frequency differences between models with and without diffusion reach several microHertz for stars with masses greater than 1.3M$_{\odot}$.

In hotter stars, the radiative accelerations may become significantly stronger than gravity for many metals, which are pushed up. The variations with depth of the radiative accelerations of specific elements combined with the selective effects of gravitational settling can lead to element accumulation or depletion in various stellar layers. In A and B-type stars, iron-peak element accumulation appear in the Z-opacity bump (located at $\simeq$200’000K). The induced opacity increase may lead to local convection \citep{Richer00,Richard01}. 

The iron-peak element accumulation in the opacity bump region can help trigger stellar pulsations, therefore improving the agreement between seismic observations and theoretical frequency spectra in many stars: e.g. in $\gamma$ Doradus, Am, SPB, $\beta$ Cephei or sdB stars \citep[see][for a detailed discussion of this subject]{Theado09}.

The effects of the radiative accelerations on the oscillation frequencies have been tested by \citet{Escobar12}. Very weak effects are observed for models with masses up to 1.28M$_{\odot}$, but significant effects are expected for more massive stars.

Progress in the understanding of the physics of early-type stars is limited by the radiative accelerations not being computed in most stellar evolutionary codes. Up to now, the Montreal stellar evolution code is the only one in which a complete, accurate, and consistent treatment of radiative accelerations has been introduced \citep{Richer00,Richard01,Turcotte00}. The price to pay for this accuracy is that the heavy and CPU-time consuming computations of atomic processes do not allow additional sophisticated treatments of macroscopic motions. In the Montreal code, turbulence is only treated as an extremely simplified process with a turbulent diffusion coefficient proportional to a parameterized power of the density. 

In this context, we introduced into the Toulouse-Geneva Evolution Code (hereafter TGEC) new computations of the radiative accelerations on heavy elements, following the semi-analytical prescription proposed by \citet{Alecian02} and \citet{LeBlanc04}. This prescription leads to fast but reasonably accurate computations, which represent a good compromise between accuracy and CPU-time consumption and allows coupling with macroscopic motions. Such a treatment of abundance variations inside early-type stars is necessary for a good understanding of these stars. 

In the present paper, we do not introduce thermohaline convection as described in \citet{Theado09} because we want to present a detailed comparison with the Montreal code in which this specific physical process is not included. Very precise tests of the results obtained by the two codes for iron accumulation inside nearly identical models with similar physics are still underway. In particular, the computations of iron fluxes seem to lead to differences in some cases, which still have to be understood. These tests are beyond the scope of the present paper and will be presented elsewhere in the near future. Here we present a first step in the comparisons, namely the detailed study of the radiative accelerations obtained with the two methods. Some abundance profiles are only shown as indicators of the results that the TGEC code is presently able to obtain.

In the following section, we explain in detail the major improvements implemented in the TGEC code. In Sect. 3, we present a comparison between the Montreal and TGEC computations for two similar stellar models. In Sect. 4, additional models are presented to illustrate the capabilities of the updated TGEC version and its application fields. Our conclusions are given in Sect. 5.

\section{New opacities and atomic diffusion computations in TGEC}
\label{implementation}
 The Toulouse-Geneva stellar evolution code is described in detail in \citet{Hui08}; however, the code has recently undergone major improvements that are reported in the following sections. The code can follow the time-dependent abundance variations of 21 species (12 elements and their main isotopes: H, $^3$He, $^4$He, $^6$Li, $^7$Li, $^9$Be, $^{10}$B, $^{12}$C, $^{13}$C, $^{14}$N, $^{15}$N, $^{16}$O, $^{17}$O, $^{18}$O, $^{20}$Ne, $^{22}$Ne, $^{24}$Mg, $^{25}$Mg, $^{26}$Mg, $^{40}$Ca, and $^{56}$Fe) in detail. The remaining metals are collected into an average species Z. 

\subsection{Opacities}
\label{opacity}
In the TGEC code, the opacities are computed using the OPCD v3.3 codes and data\footnotemark \footnotetext{The OPCD\_3.3 packages is available on the following website: http://cdsweb.u-strasbg.fr/topbase/op.html} \citep{Seaton05}. They allow computating self-consistent Rosseland opacities taking the detailed composition of the chemical mixture into account. The opacities are recalculated by considering the abundance variations at each time step and at each mesh point.

\subsection{Atomic diffusion}
\label{dif}
\subsubsection{Computational methods}
The diffusion computations are based on the Boltzmann equation for dilute collision-dominated plasma. At equilibrium, the solution of the equation is the Maxwellian distribution function. Transport properties in stars are computed considering small deviations from the Maxwellian distribution. In this framework, two different formalism have been proposed to obtain approximate solutions to the Boltzmann equations. 

The first method lies on the Chapman-Enskog theory \citep{Chapman70}, which assumes that the total distribution function of a given species can be written as a convergent series, each term of the series representing successive approximations to the distribution function. The formalism is first applied to a test element diffusing in the stellar plasma, taking the diffusion of electrons and the induced electric field into account. The computations lead to a statistical ``diffusion velocity" of the test-element with respect to the main component of the plasma. In stars, this basic component is generally hydrogen, so that the diffusion of every element is computed with respect to neutral hydrogen and/or to protons. Meanwhile, the hydrogen abundance is renormalized to recover the abundance consistency and local stellar equilibrium. For the case of helium, which has a non-negligible abundance, corrections on the diffusion velocity as proposed by \cite{Montmerle76} are added. 

The second method has been developed by \citet{Burgers69}. It is based on the Grad 13 moment approximation and the use of a Fokker-Planck collision term in the Boltzmann equation. In this approach, separate flow and heat equations for each component of a multicomponent mixture are solved simultaneously. This method provides a more convenient way for handling multicomponent gases than the Chapman-Enskog one but is heavier to apply.  
In both methods, the diffusion coefficients can be written as functions of the collisions integrals that depend on the exact nature of the interaction between colliding particules.

In TGEC, the atomic diffusion is computed following the \cite{Chapman70} formalism. 
We use the gravitational and thermal diffusion coefficients as derived by \citet{Paquette86}, who performed a detailed computation of the collision integrals for ionized elements diffusing in an ionized medium. In the case of the diffusion of neutral atoms in ions, or reverse, the polarization of the neutrals is taken into account as proposed by \citet{Michaud78}. For neutrals atoms diffusing in a neutral medium, the rigid sphere approximation is used.  
The average diffusing charge $Z_i$ of the particules is computed by solving the Saha-Boltzmann equations. The partition functions are limited to the statistical weight of fundamental levels, and no correction for high density or degenerate electrons is introduced. The set of Saha-Boltzmann equations are solved by using a Newton-Raphson iterative scheme. When the computations of radiative accelerations are added to the diffusion computations, the Chapman-Enskog method is easier to handle than Burgers'. 
 This situation helps treating cases with short time scales and rapid variations, as shown in Sect. 4. 

\subsubsection{Radiative accelerations}
\label{raddif}
The radiative accelerations on C, N, O, Ne, Mg, Ca, and Fe have been included in the TGEC code following the improved version of \citet{LeBlanc04} of the semi-analytical prescription proposed by \citet{Alecian02}. This method allows very fast computation of radiative accelerations with a reasonable accuracy.  Radiative accelerations due to bound-bound (\citealt{AlecianAl1985}, \citealt{AlecianAlAr1990a}) and bound-free \citep{AlecianAl1994} transitions are obtained using a parametric form of the radiative acceleration equation. The basic idea of this parametric method is to derive formula where the terms depending explicitly on atomic data (such as $gf$ values for instance) are separated from those depending on the stellar plasma and the abundances of the considered ion (with the aim of accounting for the saturation effects). In this framework, the radiative accelerations may be approximated by calculating a single value for each parameter found in the related equations. This is the so-called single-valued parameter (or SVP) approximation \citep{LeBlanc04}.

The interface of the SVP method with the models consists of a set of six parameters per ion, which allows 
 to estimate the radiative acceleration of each element through simple algebraic expressions \citep{LeBlanc04}, for each time step of the run of the evolution code. These parameters are determined only at the beginning of the computation through interpolation as a function of the stellar mass, inside a pre-established grid. The computation of the total radiative acceleration for a given element with SVP also requires computing the ions' relative populations. This is included in the set of the SVP numerical routines added to TGEC.

The SVP approximation may be implemented in existing codes. There are much less data to process than for detailed radiative acceleration calculations because complete and detailed monochromatic opacities for each element are not needed. A new grid of SVP-parameters, well fitted to the stellar mass range considered in this work, was computed following the procedure described in \citet{LeBlanc04}. An implementation of the SVP method was first used in 
\citet{Theado09}.

\section{Comparison with the Montreal computations}
\label{compar}

In this section, we compare the TGEC results to those obtained with the Montreal code. For this purpose we compare the internal structures and the results of the diffusion computations for two models: one computed with the Montreal code, the other computed with TGEC. The Montreal model chosen for this comparison is the 5.3D50-3, 1.7M$_{\odot}$-model presented in \citet{Richard01}. A similar model was computed with TGEC, including input physics and initial parameters as close as possible to the Montreal model. As a first step, we present comparisons of seven elements (listed in Table 1), which are especially important for A type stars.

The detailed physics of the Montreal model can be found in \citet{Richard01}, \citet{Richer00}, and \citet{Turcotte98}. The key ingredients are reported in the following sections and compared to those introduced in the TGEC model.

\subsection{Input physics}
\subsubsection{Basic input physics}
\label{classic}
Here is a list that compares the basic input physics used in the TGEC and Montreal models:

\paragraph{Equation of state: } the Montreal model uses the CEFF equation of state \citep{JCD92}.
The TGEC model is computed using the OPAL2001 \citep{Rogers02} equation.

\paragraph{Opacities: } in both models, the opacities are computed at every point in the star and at each evolution time step. The Montreal computations take the local abundance of 21 chemical elements into account \citep[see Table 1 of][]{Turcotte98} using the OPAL monochromatic data. In TGEC, the opacities are computed as described in Sect. \ref{opacity} considering the detailed abundance of the elements listed in Table \ref{initabund}.

\paragraph{Nuclear reactions: } the Montreal code uses the nuclear energy generation routine of \cite{Bahcall92}. The nuclear reaction rates used in TGEC are from the analytical formulae of the NACRE compilation \citep{Angulo99}.

\paragraph{Convection: } in both Montreal and TGEC models, the delimitation of the convective zones is based on the Schwarzschild criterion. The energy flux in the convection zones is computed following the B\"ohm-Vitense mixing length formalism. In the Montreal model, convective zone mixing is modeled as a diffusion process, described by a diffusion coefficient $D_{mix}$ (cf. Sect. \ref{transportmacro}). A mixing length approximation is used for $D_{mix}$, which always leads to high $D_{mix}$ values and very homogeneous convective zones. In TGEC, convective zones are assumed to be instantaneously homogenized. The mixing length parameter $\alpha$ is taken to be equal to 1.68 in both the TGEC and Montreal models.

\paragraph{Initial chemical mixture: } the initial mixture used in the Montreal model is given in Table 1 of \citet{Turcotte98}. The initial composition introduced in the TGEC model is given in Table \ref{initabund} of this paper: the chemical elements followed in both models are introduced with the same initial abundance.

\begin{table}
  \caption[]{Initial chemical composition (in mass fraction) for models presented in Section 3.}
  \label{initabund}
  \begin{tabular}{c c}
            \noalign{\smallskip}
            \hline
            \noalign{\smallskip}
            H & 0.70\\
            ($^3$He+$^4$He) & 0.28\\
            $^{12}$C & $3.466 \times 10^{-3}$\\
            $^{14}$N & $1.063 \times 10^{-3}$ \\
            $^{16}$O & $9.645 \times 10^{-3}$\\
            $^{40}$Ca & $7.469 \times 10^{-5}$\\
            $^{56}$Fe & $1.436 \times 10^{-3}$ \\
            \noalign{\smallskip}
            \hline
            \noalign{\smallskip}
  \end{tabular}
\end{table}

\subsubsection{Atomic diffusion}
In the Montreal code, atomic diffusion is computed as described in \citet{Richard01} and \citet{Turcotte98}. The Montreal calculations take the time-dependent variations of 28 species (including isotopes) into account: H, $^3$He, $^4$He,$^6$Li, $^7$Li, $^9$Be, $^{10}$B, $^{11}$B, $^{12}$C, $^{13}$C, N, O, Ne, Na, Mg, Al, Si, P, S, Cl, Ar, K, Ca, Ti, Cr, Mn, Fe, and Ni. Monochromatic OPAL data are used to calculate the Rosseland opacities and the radiative accelerations at each evolution time step. The abundances are updated at every iteration over the stellar structure. The diffusion coefficients and velocities are determined by solving the Burgers's flow equations for ionized gases \citep{Burgers69} for all diffusing elements. The collisions integrals are from \citet{Paquette86}.

As a first step, we include, in the TGEC model, the atomic diffusion (including radiative accelerations) of seven elements (eight species including isotopes), especially important for A type stars: H, $^3$He, $^4$He, $^{12}$C, $^{14}$N, $^{16}$O, $^{40}$Ca, and $^{56}$Fe. The atomic diffusion is computed as described in Sect. \ref{dif}. The diffusion velocities are computed following the \citet{Chapman70} formalism in the test-atom approximation, and the diffusion coefficients are from \citet{Paquette86}.

\subsubsection{Macroscopic transport of chemical elements}
\label{transportmacro}
\textit{Macroscopic transport in the Montreal code.} \\
In the Montreal code, macroscopic transport processes, including (semi)convection and turbulent mixing, are modeled as diffusion processes by adding a pure diffusion term to each element's diffusion velocity $V_{pi}$: 
\begin{equation}
V_i(macro)=-(D_T+D_{mix}) \frac{\partial \ln X_i}{\partial r}
\label{eqvdif2}
\end{equation}
where $D_T$ and $D_{mix}$ are turbulent diffusion coefficients with $D_{mix}$ representing the effects of convective and semiconvective motions and $D_T$ parametrizing turbulent transport. The variable $X_i$ represents the mass fraction of species $i$.

In radiative layers $D_{mix}=0$, while in convective zones it is computed using the mixing length theory (as stated in Sect. \ref{classic}). In semiconvection zones, $\rm D_{mix}$ is assumed proportional to $(\nabla-\nabla_{ad})/(\nabla_L-\nabla)$, where $\nabla$ and $\nabla_{ad}$ stands for the local and adiabatic logarithmic temperature gradients and $\nabla_L$ is a function of $\nabla_{\mu} = d \ln \mu / d \ln P$. We refer the reader to Sect. 2.2 of \citet{Richard01} for a more detailed description of the treatment of convection.

\begin{figure}
\includegraphics[width=0.5\textwidth,bb = 18 430 592 718,clip=true]{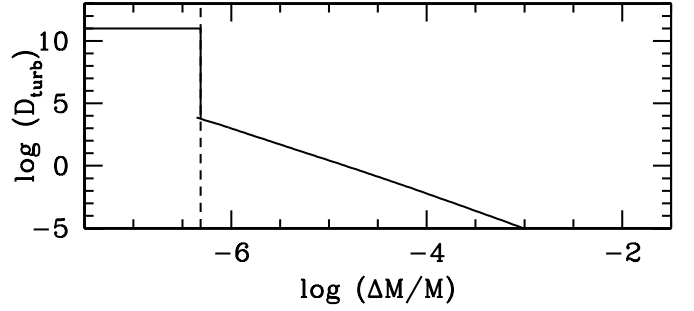}%
\caption{Turbulent diffusion coefficient in the 170Myr, 1.7M$_{\odot}$-model computed with TGEC. $D_{turb}$ is constant from the surface down to $\log T=5.3$. Just below this region, $D_{turb}$ is by definition 50 times larger than the He atomic  diffusion coefficient (see Eq. \ref{eqdt}). The vertical dashed line locates the region where $\log T=5.3$.}
\label{figdt}
\end{figure}

In the 5.3D50-3-model, the turbulent transport is chosen strong enough to guarantee, during the whole evolution period considered, a complete mixing throughout the region between the surface and the point where $\log T_0=5.3$ \citep[see Fig. 2 of][]{Richard01}. This mixing mimics the effects of a Fe convection zone except that it is imposed throughout evolution. Below $\log T_0=5.3$, the diffusion coefficient $\rm D_T$ obeys the following algebraic dependence on density \citep[Eq. 1. of][]{Richer00}:
\begin{equation}
\label{eqdt}
D_T=\omega D(He_0) \biggl(\frac{\rho_0}{\rho} \biggr)^n
\end{equation}
where $\rho_0=\rho(T_0)$, $\omega=50$, $n$=3 and 
\begin{equation}
D(He)=3.3 \times 10^{-15} T^{2.5} / [4 \rho \ln (1+1.125\times 10^{-16}T^3/\rho)].
\end{equation}
\textit{Macroscopic transport in the TGEC code.} \\
In the TGEC code, the macroscopic motions of a given species $i$ are parametrized, as in the Montreal code, by including a turbulent diffusion term to the diffusion velocity of the considered species:
\begin{equation}
V_i(turb)=-D_{T} \frac{\partial \ln X_i}{\partial r}.
\label{eqvdif3}
\end{equation}
The TGEC code does not include a $D_{mix}$ term, since convection is treated as an instantaneous dilution.
The diffusion coefficient $D_{T}$ is chosen to mimic the mixing $(D_T+D_{mix})$ introduced in the 5.3D50-3 Montreal model. Figure \ref{figdt} represents this diffusion coefficient inside our model, it may be compared to Fig. 2 of \cite{Richard01}. From the surface down to $\log T=5.3$, $D_T$ is chosen to homogenize the stellar material, below this region $D_{turb}$ is computed using the same expression as introduced in the Montreal code (i.e. following Eq. \ref{eqdt}). 

\subsection{Results}
The Montreal and TGEC codes compute stellar evolution from pre-main sequence up to the subgiant branch. They were assumed to be homogeneous on the pre-main sequence, atomic diffusion, and turbulent transport were introduced at the beginning of the main-sequence phase. 

\begin{figure}
\includegraphics[width=0.5\textwidth]{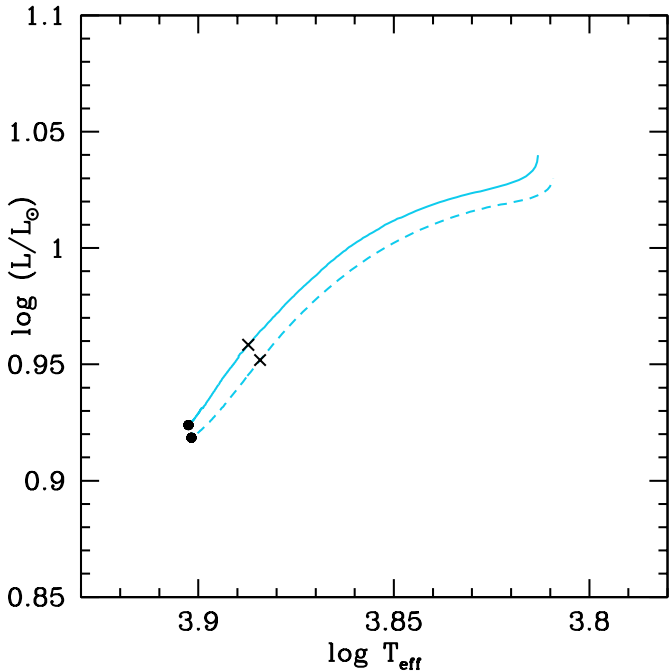}%
\caption{Evolutionary tracks of two 1.7M$_{\odot}$ sequences computed with TGEC or with the Montreal code. The dashed curve represents the 5.3D50-3 model presented in \citet{Richard01} (and computed with the Montreal code). The solid curve represents a model computed with TGEC and including similar input physic. The black dots and crosses respectively represent the 30 and 400Myr-models.}
\label{diaghr2}
\end{figure}

Figure \ref{diaghr2} compares the evolutionary tracks of the two models (for clarity the pre-main-sequence evolution is not shown). In spite of different treatments of some physical ingredients (e.g. equation of state, opacities, nuclear reactions, diffusion), the two tracks appear relatively close to each other.

\begin{figure*}
\includegraphics[width=0.66\textwidth,bb= 18 431 580 703,clip=true]{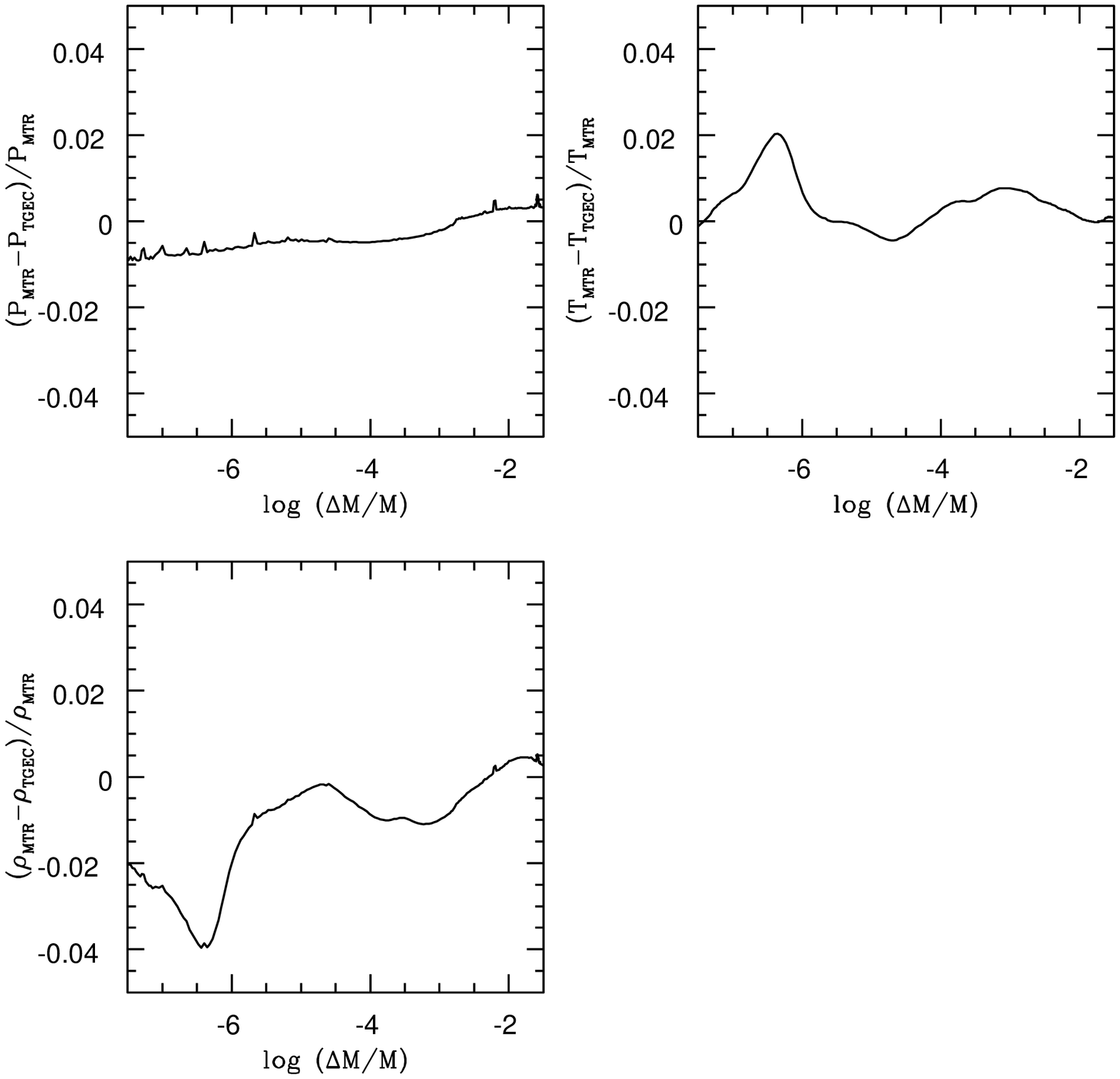}%
\includegraphics[width=0.33\textwidth,bb= 18 159 299 431,clip=true]{fig3.eps}
\caption{Comparison of the internal structures (pressure, temperature and density) of the two 1.7M$_{\odot}$-models represented in  Fig. \ref{diaghr2}. The comparison is shown at 30Myrs. MTR stands for the model computed with the Montreal stellar evolution code while TGEC stands for the model computed with the Toulouse Geneva stellar evolution code.}
\label{internalstruc}
\end{figure*}

We first give a detailed comparison between the Montreal and TGEC calculations for the two 30Myr models. Figure \ref{internalstruc} compares the internal structure of these two models. The relative differences in pressure, temperature, and density never exceed 6\% (and even less for P and $\rho$). The maximum relative differences are observed near $ \log (\Delta M/M) \simeq -6.3$, which corresponds to the base of the potential iron convective zone.

\begin{figure*}
\includegraphics[width=0.017\textwidth,bb= 30 144 60 718]{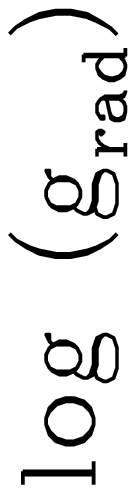}%
\includegraphics[width=0.32\textwidth]{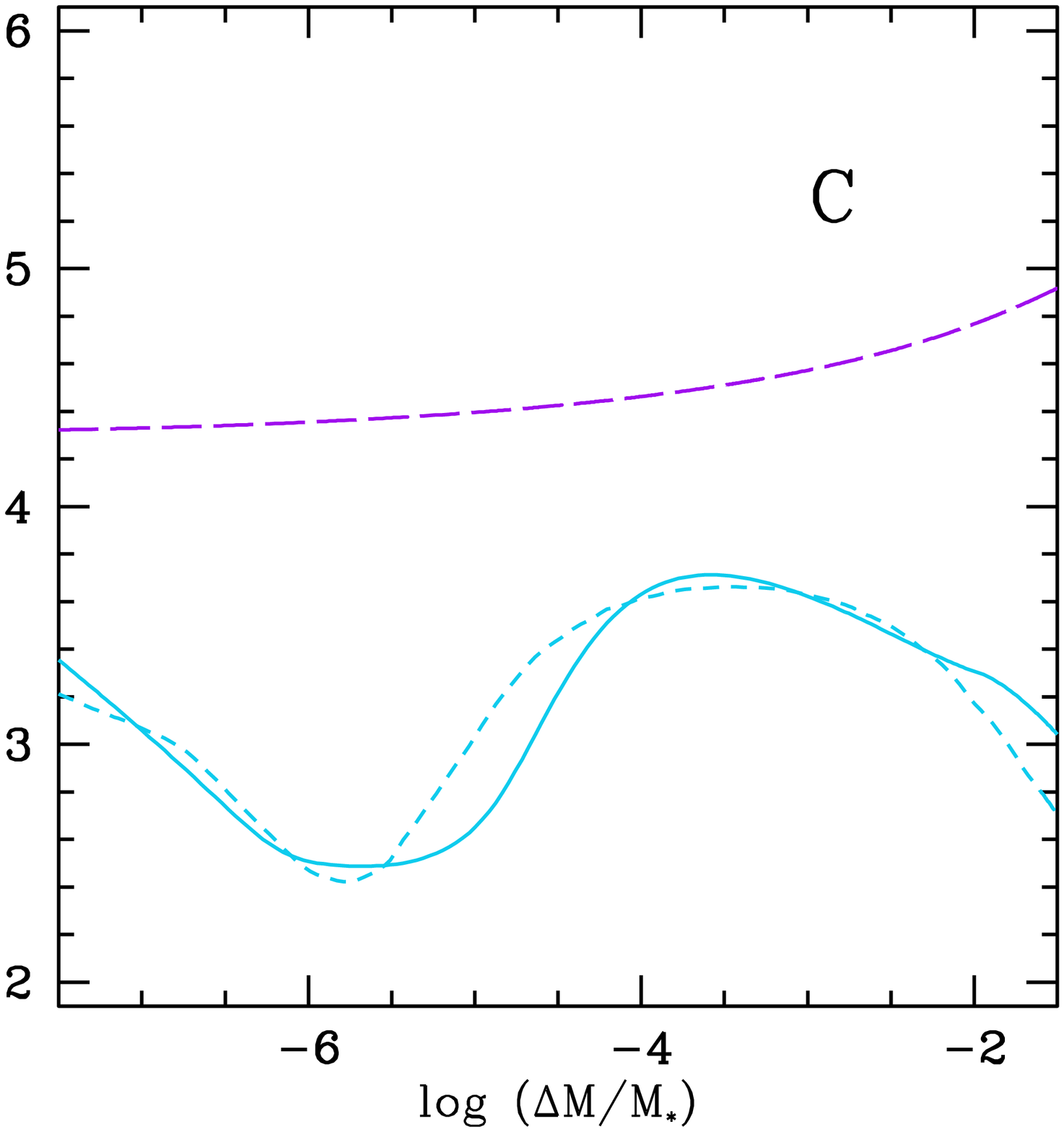}%
\includegraphics[width=0.32\textwidth]{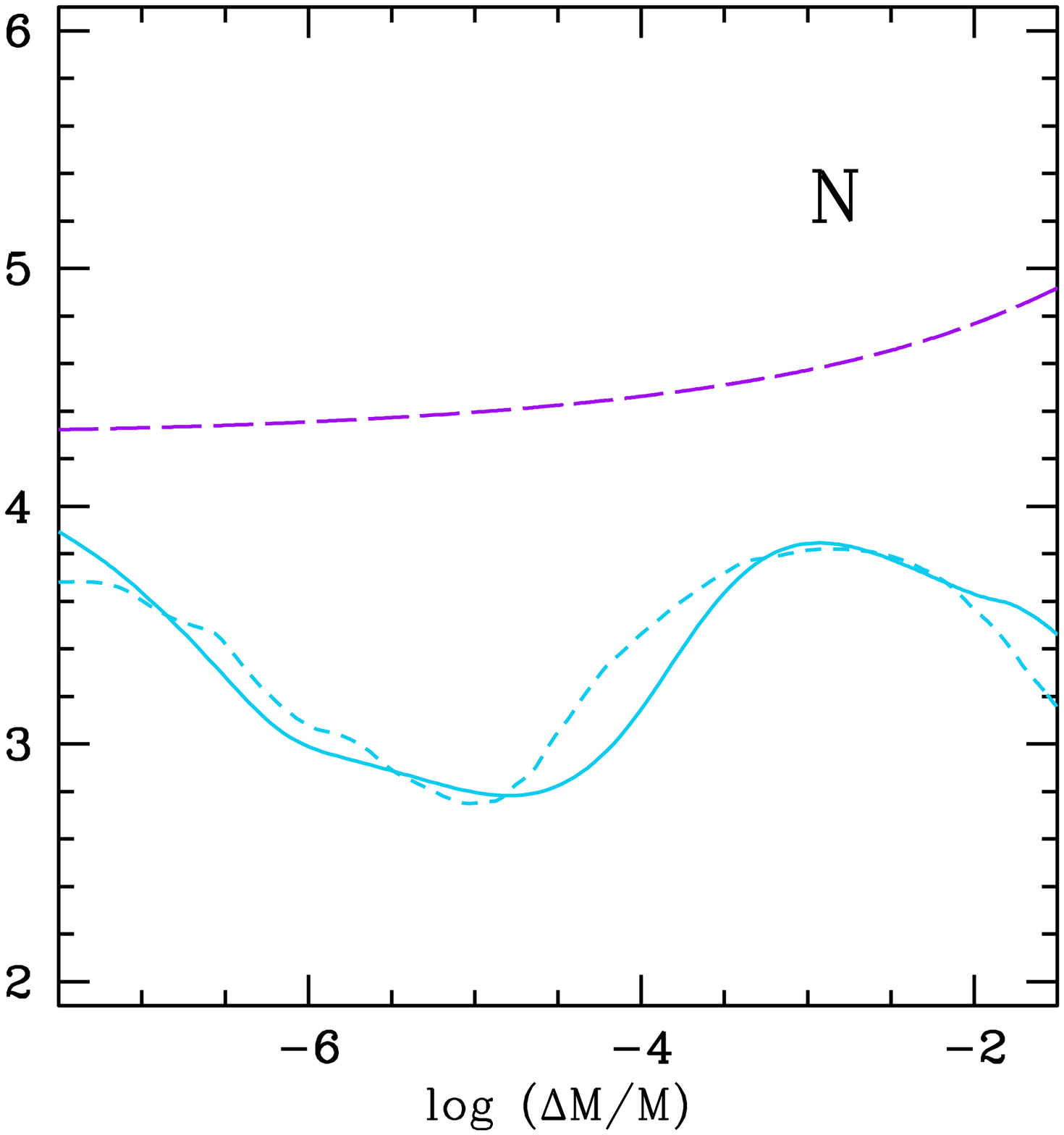}%
\includegraphics[width=0.32\textwidth]{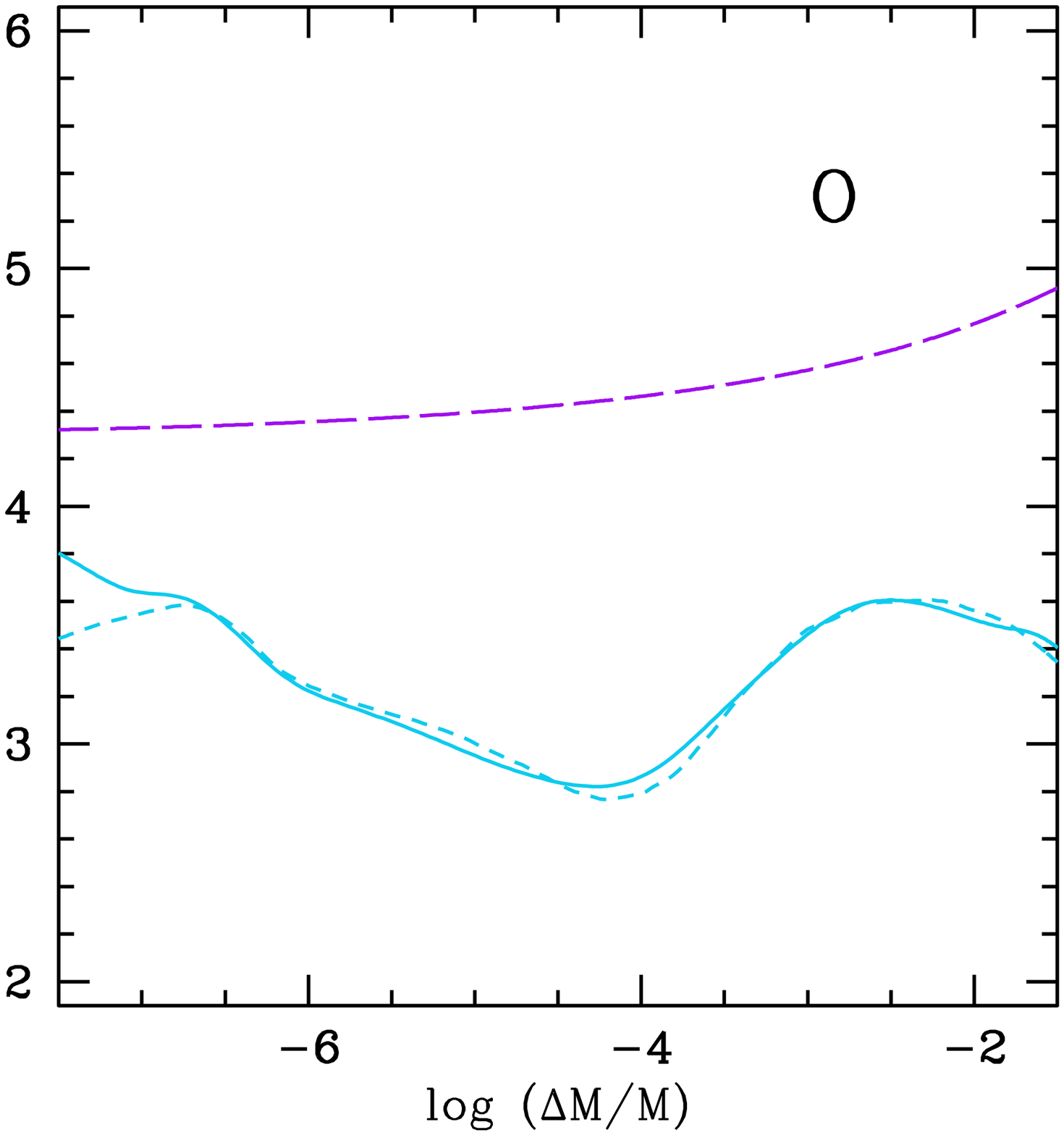}
\includegraphics[width=0.017\textwidth,bb= 30 144 60 718]{fig4a.eps}%
\includegraphics[width=0.32\textwidth]{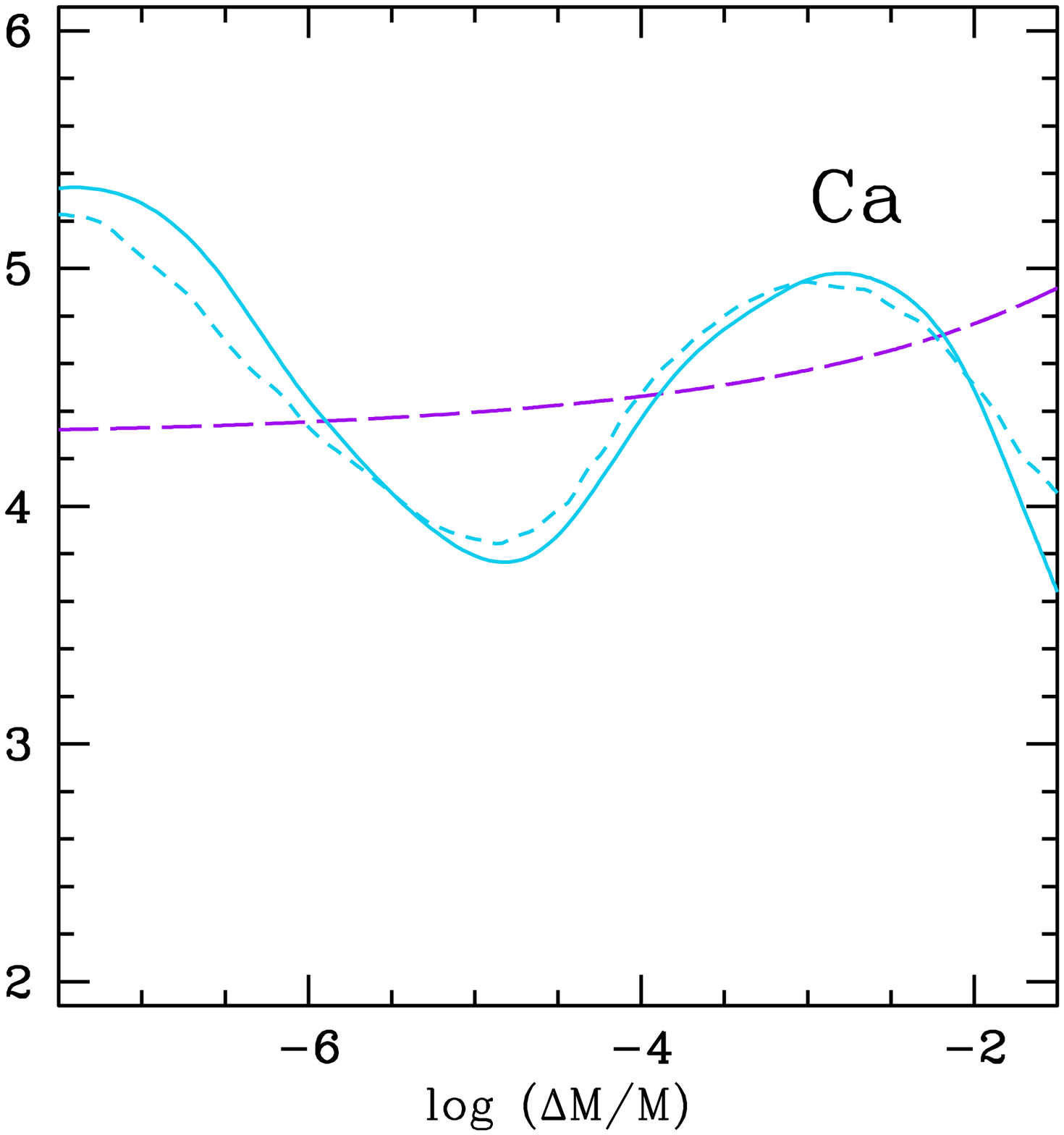}%
\includegraphics[width=0.32\textwidth]{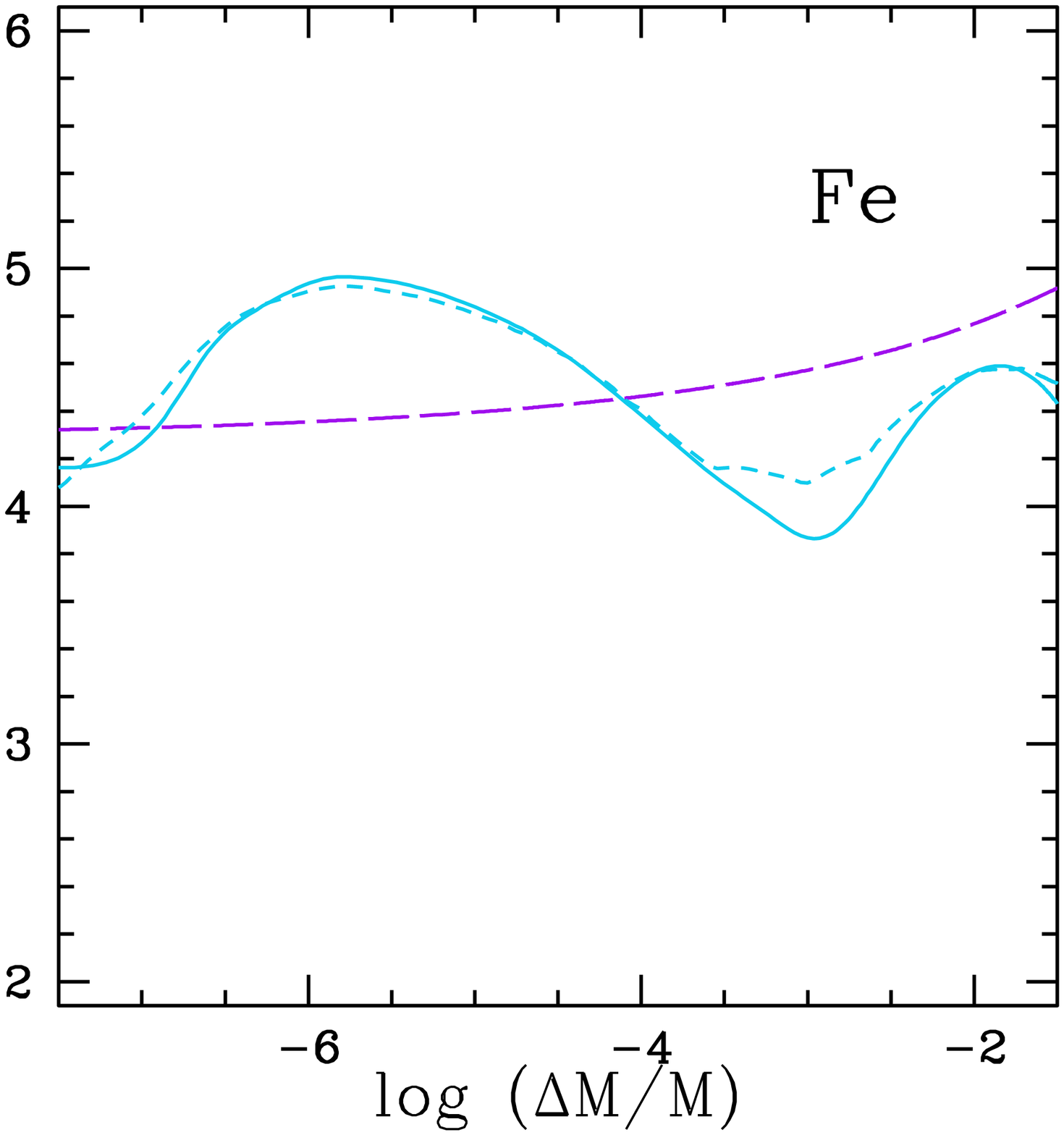}%
\caption{Radiative accelerations (in cgs unit) at 30Myr in the TGEC and Montrealmodels represented in Fig. \ref{diaghr2}. The solid curves represents the accelerations in the TGEC model, while the dashed curve represent those in the Montreal model. The long-dashed curve represents the local gravity in the TGEC model. (The local gravity in the Montreal model is not represented as it is indistinguishable from that of the TGEC model.)}
\label{grad}
\end{figure*}

Similarly to Fig. 7 of \citet{Richard01}, Fig. \ref{grad} displays the radiative accelerations and the local gravity of the two models. The discrepancy between g$_{MTR}$ and g$_{TGEC}$ remains smaller than around 0.3 dex except for O in the upper layers (in the convection zone). We notice that the average accuracy of the SVP method is estimated to $\pm$0.1 dex \citep{Alecian02} with respect to the detailed computations of \cite{Seaton97}. However, the SVP approximation is less accurate for light elements (lighter than O) than for heavy ones because this method uses parameters determined at the position of the maximum of each ion relative population. Because there are fewer ionization states for light elements, the gap in temperature between these maxima is larger for them. The comparison of Fig. \ref{grad} is fairly satisfactory, since radiative accelerations are computed in completely different ways and use different atomic and opacity databases.

\begin{figure*}
\includegraphics[width=0.04\textwidth,bb= 10 44 60 400,clip=false]{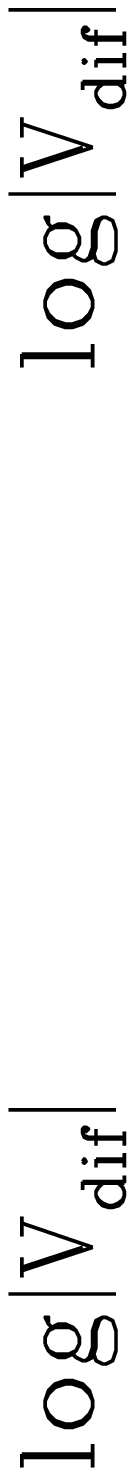}%
\includegraphics[width=0.32\textwidth,bb= 300 144 592 718,clip=true]{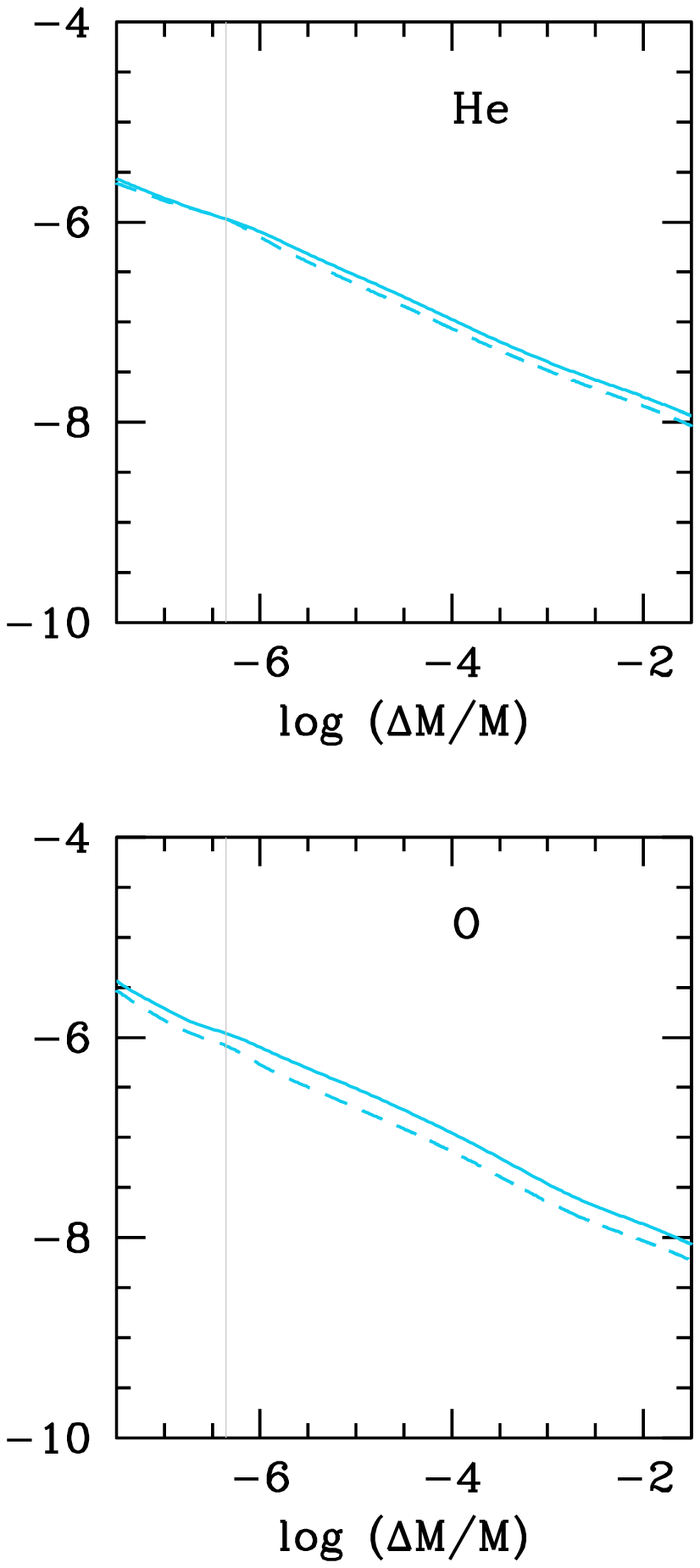}%
\includegraphics[width=0.32\textwidth,bb= 300 144 592 718,clip=true]{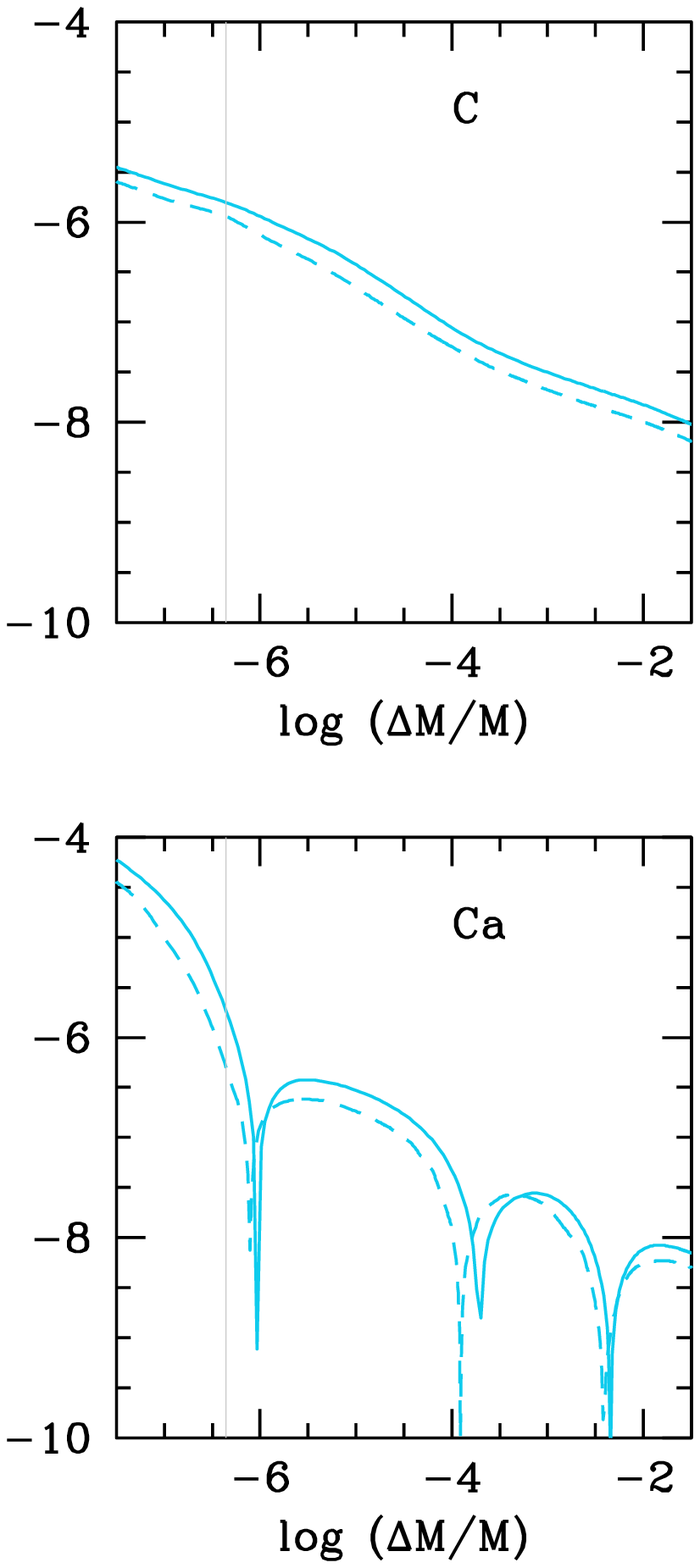}%
\includegraphics[width=0.32\textwidth,bb= 300 144 592 718,clip=true]{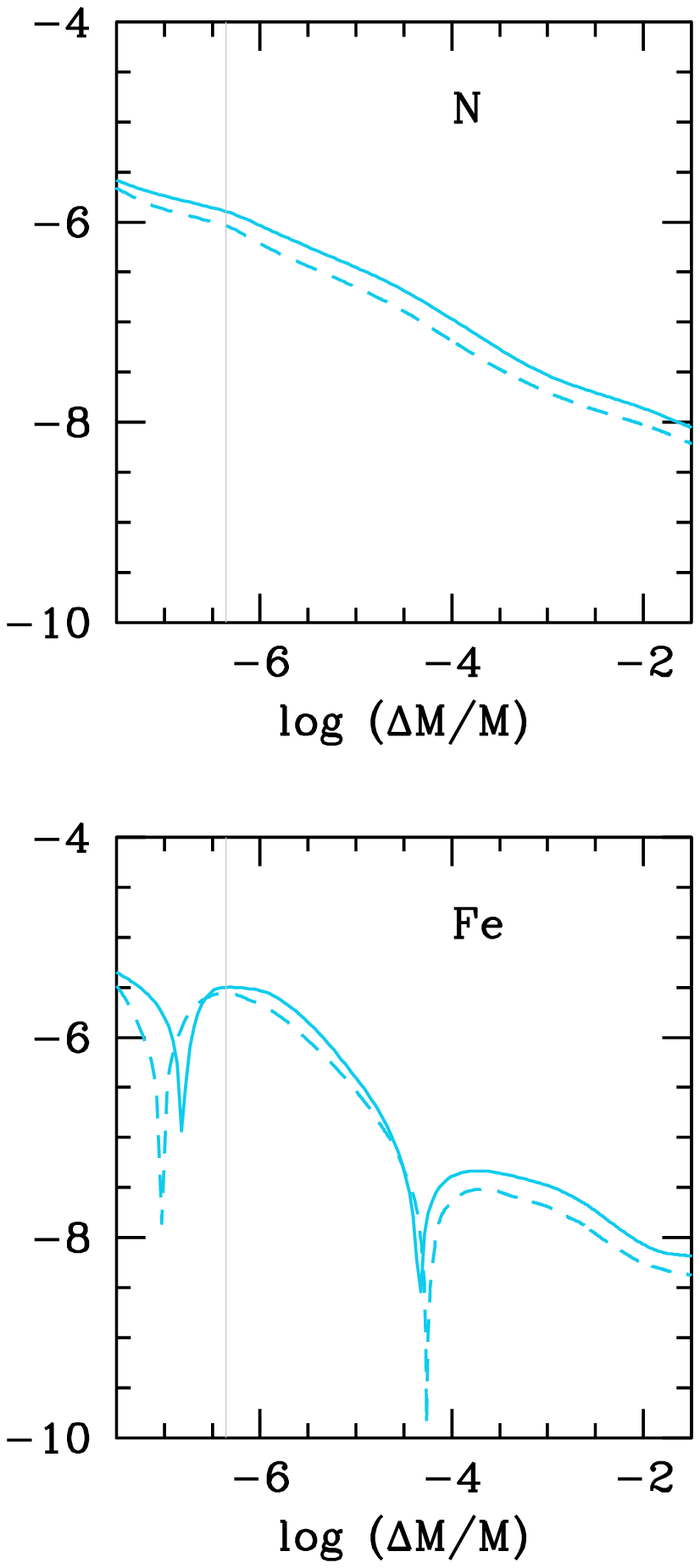}
\caption{Diffusion velocities (in cgs units) below the surface convective zone in the Montreal (dashed curves) and TGEC (solid curves) 30Myr-models represented in Fig. \ref{grad}. The vertical lines locate the bottom of the completely mixed region. The peaks observed for Ca and Fe correspond to sign changes in the diffusion velocities: the Ca diffusion velocity is positive for $\log (\Delta M/M) \lesssim -6$ and $-3.5  \lesssim \log (\Delta M/M) \lesssim -2.5$. The Fe diffusion velocity is positive for $-7 \lesssim \log (\Delta M/M) \lesssim -4.5$.}
\label{vdif}
\end{figure*}

\begin{figure*}
\includegraphics[width=0.017\textwidth,bb= 30 144 60 718]{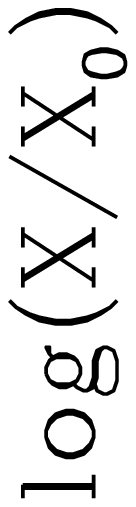}%
\includegraphics[width=0.32\textwidth]{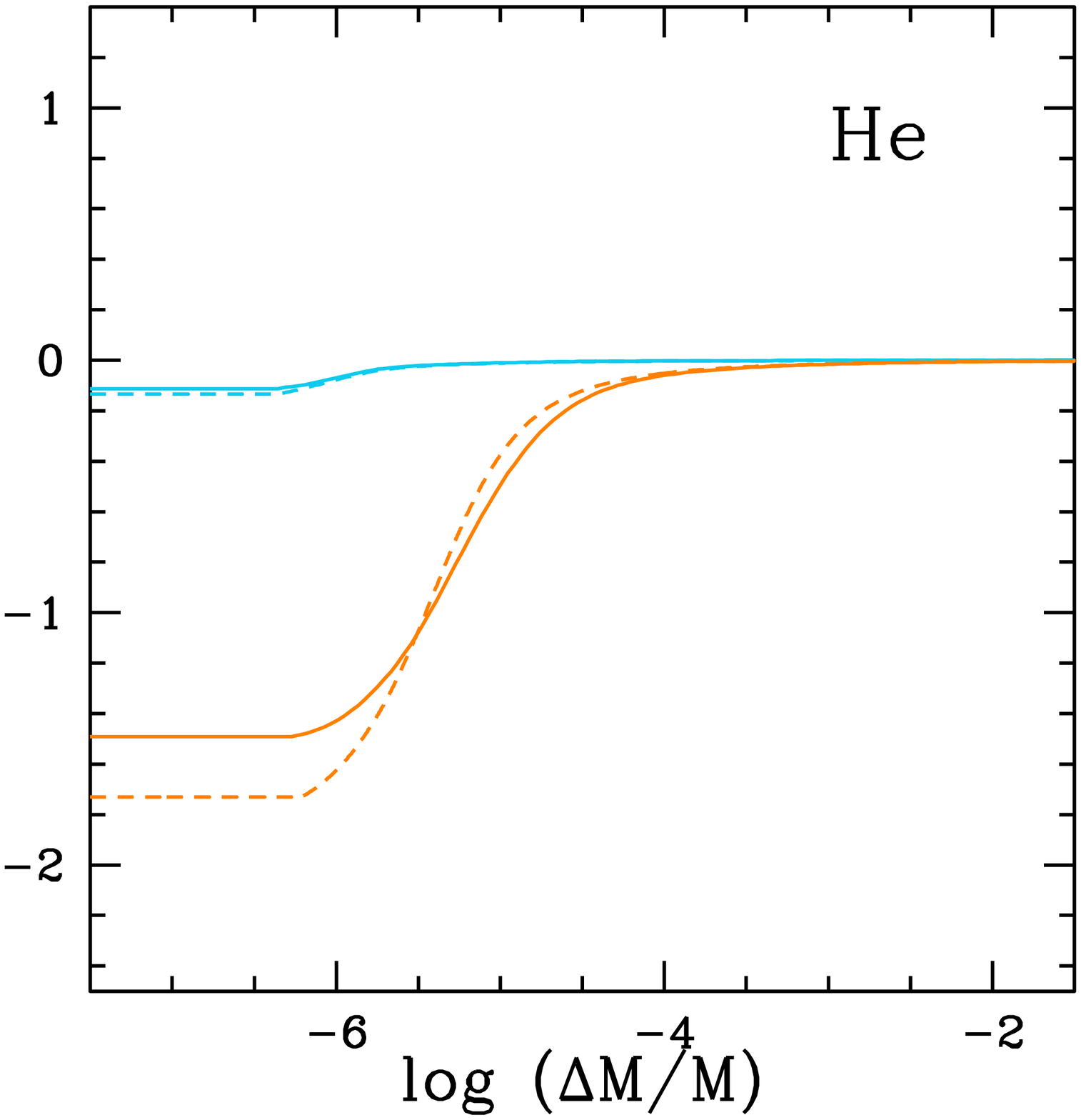}%
\includegraphics[width=0.32\textwidth]{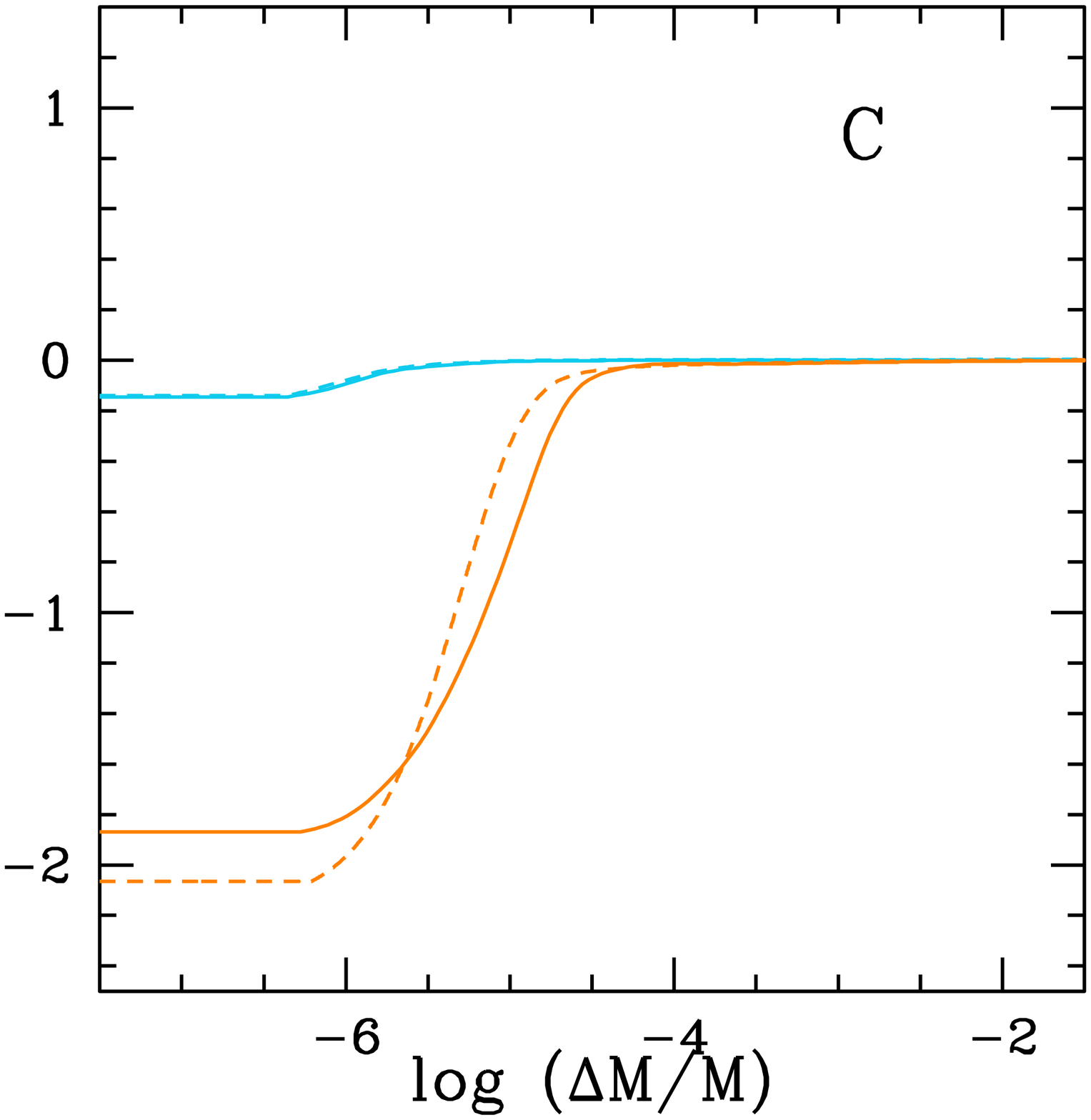}%
\includegraphics[width=0.32\textwidth]{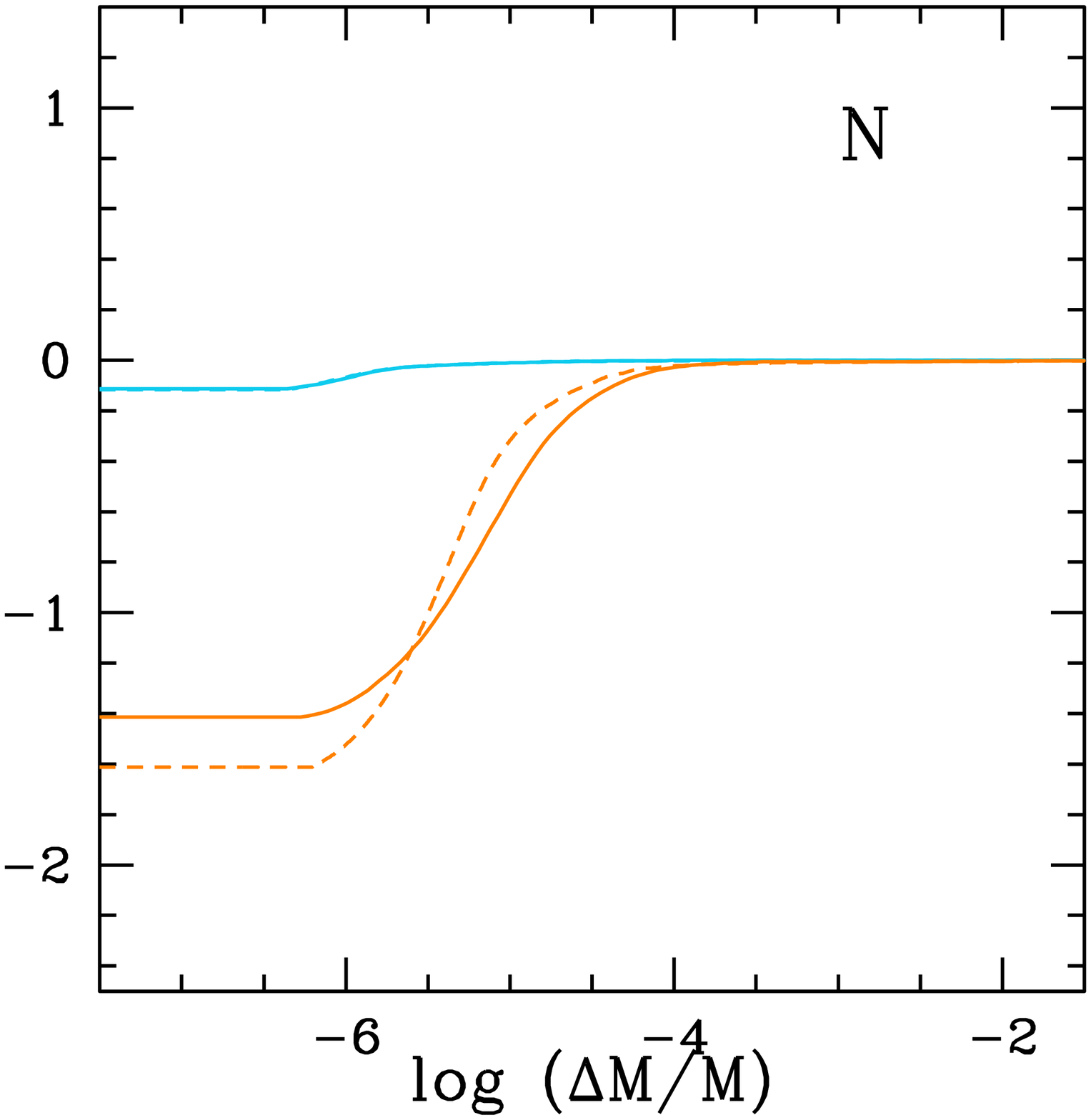}
\includegraphics[width=0.017\textwidth,bb= 30 144 60 718]{fig6a.eps}%
\includegraphics[width=0.32\textwidth]{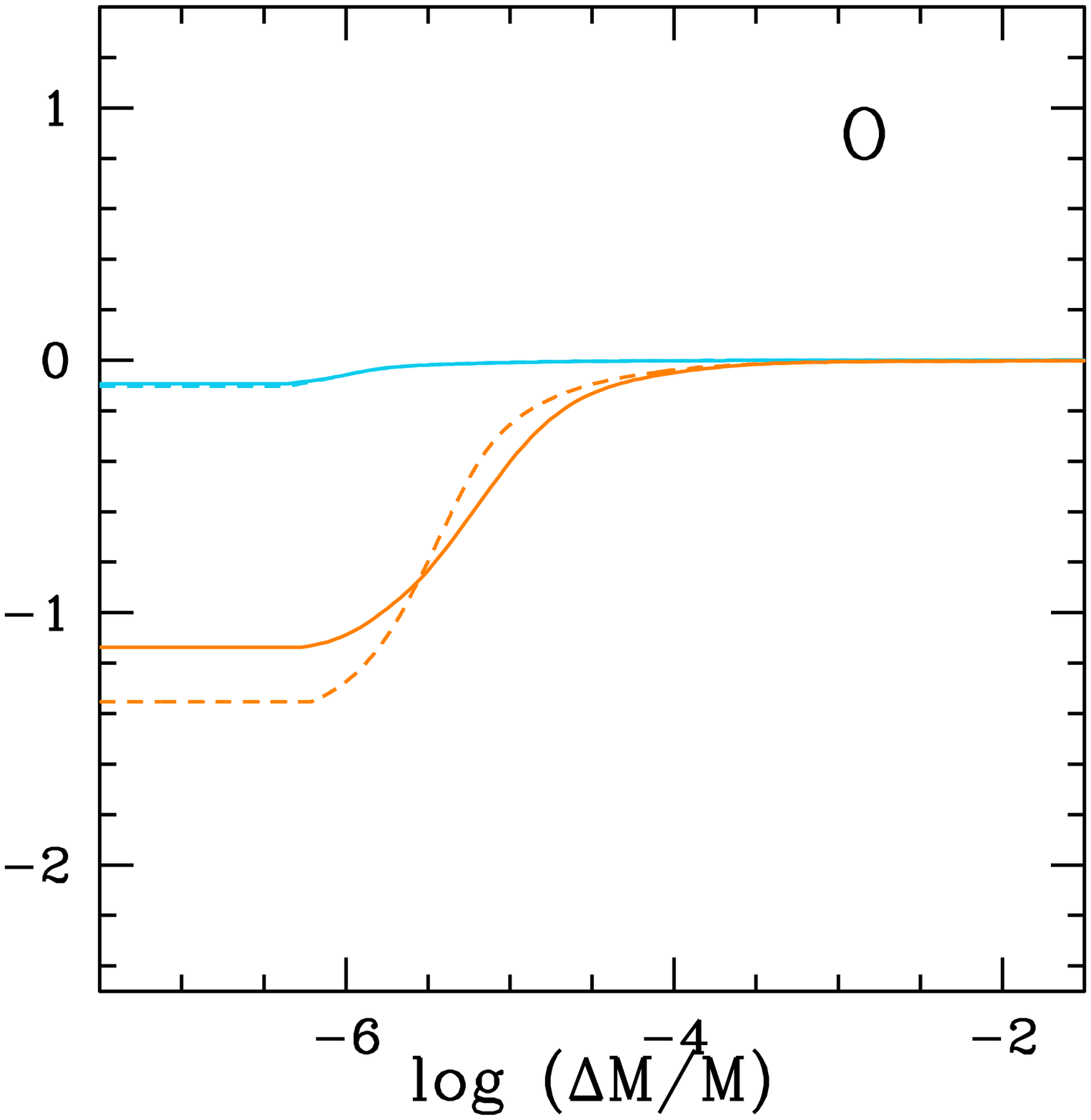}%
\includegraphics[width=0.32\textwidth]{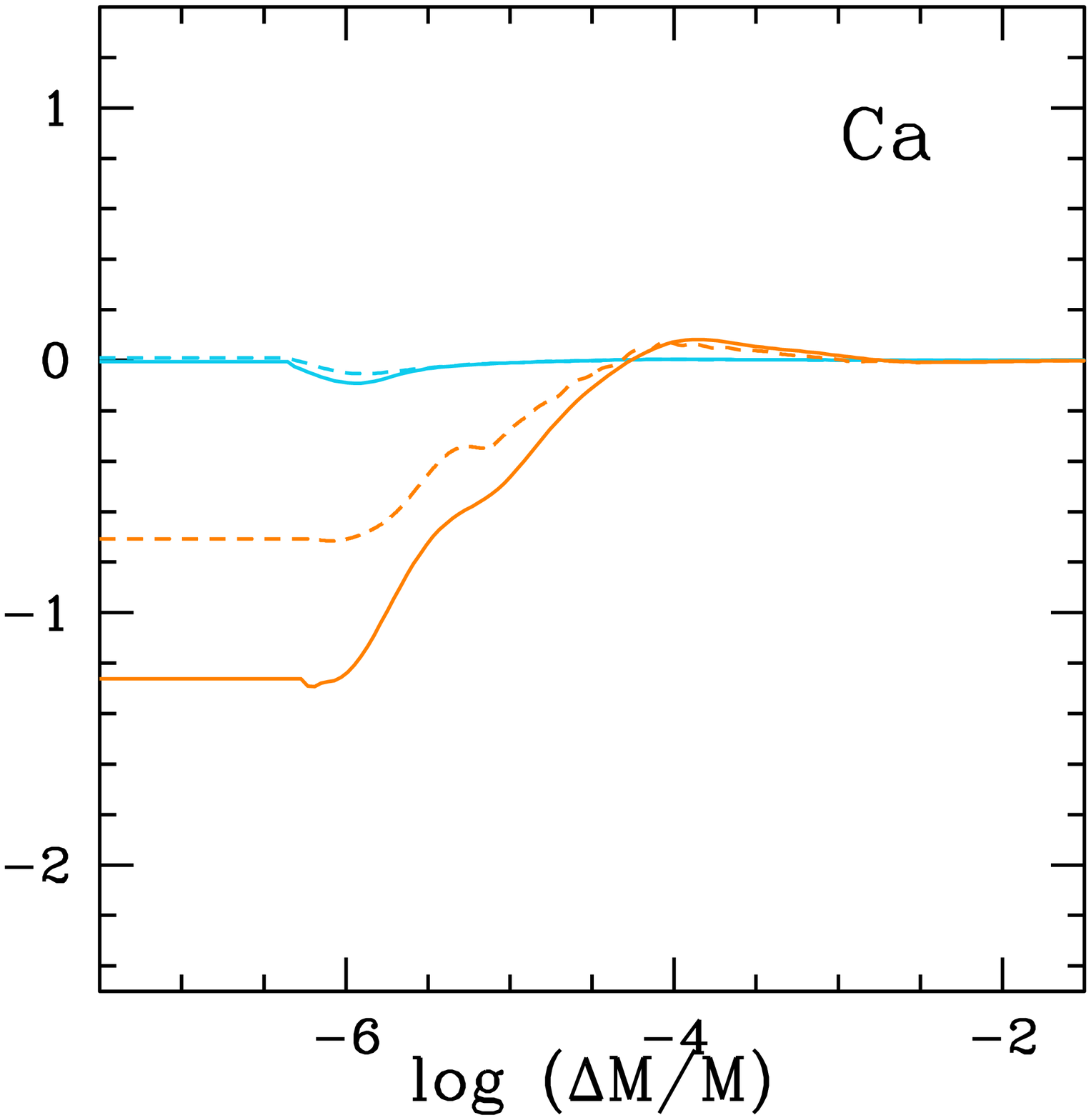}%
\includegraphics[width=0.32\textwidth]{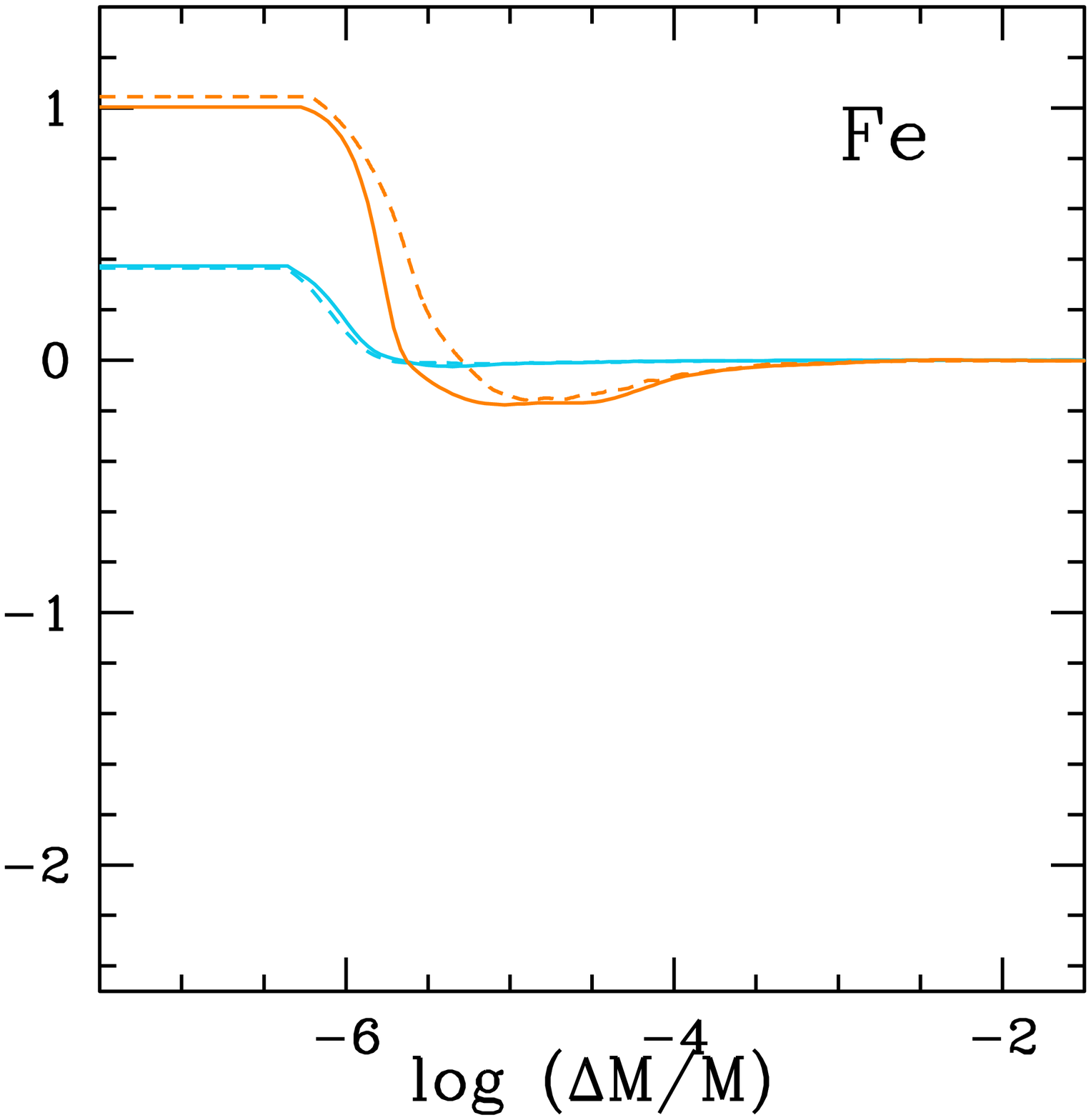}
\caption{Abundance profiles in the TGEC (solid curves) and Montreal (dashed curves) models shown in Fig. \ref{diaghr2} as a function of the outer mass fraction. $\log(X/X_0)$ represents the ratio between the current element mass fraction and its initial values (at t=0). The blue and orange curves present the abundances at 30 Myr and 400 Myr after the ZAMS, respectively. }
\label{abond}
\end{figure*}

Figure \ref{vdif} presents the diffusion velocities of several elements for the same models. Despite the different atomic diffusion prescriptions used, the diffusion velocities are quite close in most of the star.
Figure \ref{abond} shows the abundance profiles in the two 30 and 400Myr-models. At 30 Myr the profiles are quite close for all the species followed in TGEC. However, at 400 Myr discrepancies in the abundance profiles are observed. These differences are small for most elements (He, C, N, O, and Fe) except for Ca.
As shown in Fig. \ref{diaghr2}, the positions of the two 400Myr models in the HR diagram are slightly different, and are in particular more distant from each other than the 30Myr models. As a consequence, their internal structure is also expected to be different. In this context, it seems difficult to disentangle the discrepancies in the models due to structure variations from those caused specifically by the difference in diffusion calculations.

\begin{table*}
\caption{Characteristics of the models.}
\medskip
\centering
\begin{tabular}{cccccccc}
\hline\noalign{\smallskip}
   & Age (Myr) & $X_c$ & $|\Delta X_c|/X_c$ & $L/L_{\odot}$ & $|\Delta L|/L$ & $T_{eff}$ (K) & $|\Delta T_{eff}|/T_{eff}$  \\
\hline\noalign{\smallskip}
M$_{MTR}$ & 30.2 & 0.6938 & 0.03\% & 8.289 & 1.25\% & 7975.66 & 0.18\% \\
M$_{TGEC}$ & 30.2 & 0.6940 &        & 8.393 &        & 7989.90 &        \\
\hline\noalign{\smallskip}
M$_{MTR}$ & 400 & 0.5778 & 0.81 \% & 8.951 & 1.51\% & 7659.49 & 0.70\% \\
M$_{TGEC}$ & 400 & 0.5731 &         & 9.086 &        & 7713.35 &        \\
\noalign{\smallskip}
\hline
\end{tabular}
\label{tab}
\end{table*}

\section{Application field of the TGEC code}
In this section, we propose to demonstrate the abilities of the TGEC code and its application field. Models of A-F stars including atomic diffusion and minimal mixing (i.e. only a mild mixing below the convective zones to avoid steep and unrealistic composition gradients at the transition between radiative and convective regions) have already been presented in \citet{Theado09}. These models, which were evolved until the end of the main-sequence phase, have shown the ability of the code to compute complete evolutionary tracks including radiative levitation effects in the presence of minimal mixing.

In this paper, we propose another test of the abilities of the code to manage rapid variations in the chemical composition and the thermal structure. For this purpose we present a set of models computed with TGEC with masses ranging from 1.5 to 2.5M$_{\odot}$. Like the models presented in \cite{Theado09}, they include atomic diffusion and a mild mixing below the convective zones. In the present models, the iron convective zone, when it appears, is assumed to be connected and completely mixed with the H/He convective envelope.
These models were evolved from pre-main sequence up to hydrogen-core exhaustion. They were assumed to be chemically homogeneous on the pre-main sequence. Atomic diffusion was introduced at the beginning of the main-sequence phase. Because the present computations are done to test the TGEC code, we did not include here the thermohaline convection that must be added for comparison with real stars \citep{Theado09}.

\label{computations1}
\subsection{Input physics}
As previously described, our models were computed using the OPAL2001 \citep{Rogers02} equation of state. The nuclear reactions were from the NACRE compilation \citep{Angulo99}, and the opacities were computed as described in Sect. \ref{opacity}. Atomic diffusion (including the radiative forces) was introduced as described in Sect. \ref{dif}, for the following elements: H, $^3$He, $^4$He, $^{12}$C, $^{14}$N,$^{16}$ O, $^{40}$Ca, and $^{56}$ Fe.
The initial metal mixture was the solar mixture of \cite{Grevesse93}. The initial mass fractions for the diffusing species are given in Table \ref{table1}. Convection was computed using the mixing length theory with a mixing length parameter $\alpha=1.8$. The HI and HeII convective zones were assumed to be connected by overshooting and mixed together. As stated previously, the iron convective region that may appear during main-sequence evolution was also assumed connected and completely mixed with the surface convective region.
\begin{table}
  \caption[]{Initial chemical composition (in mass fraction) for models presented in Section 4.}
  \label{table1}
  \begin{tabular}{c c}
            \noalign{\smallskip}
            \hline
            \noalign{\smallskip}
            H & 0.7112\\
            ($^{3}$He+$^{4}$He) & 0.2714 \\
            $^{12}$C & 0.2981447E-02\\
            $^{14}$N & 0.9218931E-03 \\
            $^{16}$O & 0.8375229E-02\\
            $^{40}$Ca & 0.6283632E-04\\
            $^{56}$Fe & 0.1148674E-02 \\
            \noalign{\smallskip}
            \hline
            \noalign{\smallskip}
  \end{tabular}
\end{table}

\begin{figure}
\includegraphics[width=0.5\textwidth,bb = 18 430 592 718,clip=true]{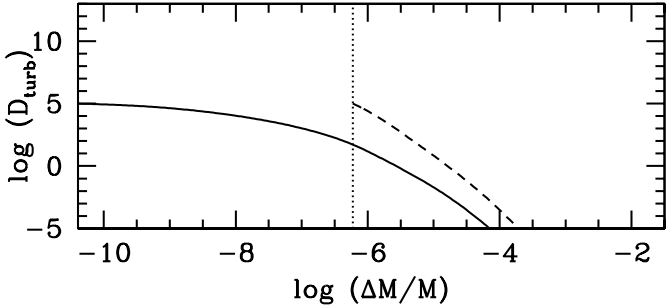}
\caption{Turbulent diffusion coefficient (according to Eq. \ref{eq5}) in a 1.7M$_{\odot}$-model computed with TGEC at two evolutionary stages. The solid line represents the diffusion coefficient below the surface convective zone (due to the ionization of H and He) at 170Myr, the dashed line represents the diffusion coefficient below the iron convective zone (assumed connected with the H and He ones) at 588Myrs, the vertical dotted line represents the base of the iron convective zone.}
\label{dtacho}
\end{figure}

To avoid the appearance of sharp and nonphysical abundance gradients at the transition between radiative and convective regions, we introduced mild mixing below the surface convective zone. This mixing was modeled as a diffusion process \citep{SchatzmanSc1969} as described in Eq. \ref{eqvdif3} with

\begin{equation}
D_{T}=D_{czb} \exp \left( \ln 2 \frac{r-r_{czb}}{\Delta} \right),
\label{eq5}
\end{equation}
where $D_{czb}$ and $r_{czb}$ are the value of $D _{T}$ and the value of the radius at the base of the convective zone, respectively and $\Delta$ is the half width of the mixing region, $D_{czb}$ and $\Delta$ are free parameters chosen to produce a mild mixing to a small extent. The value of $D _{czb}$ is taken as equal to $2 \times 10^5$cm$^2$.s$^{-1}$, and $\Delta$ is fixed to 0.5\% of the radius below the surface convective region. Figure \ref{dtacho} presents this diffusion coefficient below the surface convective zone of 1.7M$_{\odot}$ models at two evolutionary stages. At 170 Myr, the surface convective region includes the H and He convective zones (there is no Fe convective zone), and at 588 Myrs the external convective zone is composed of the H, He, and Fe convective zones assumed connected and mixed.

The physics introduced in these models is close to that of the r30-3M, 1.5M$_{\odot}$-model presented in \citet{Richard01}. Like our models, the r30-3M model includes mild mixing below the He convective zone. The turbulent diffusion coefficient is chosen as large as $\rm 10^4 cm.s^{-1}$ at the base of the convective zone. It then rapidly decreases with depth, varying like $\rho^{-3}$. Turbulent transport is chosen to be large enough to guarantee complete mixing between the Fe and He convective zones whenever the Fe convective zone appears. The initial metal mixture and the mixing length parameter are, however, different in the Montreal and TGEC models. For CPU-time consuming reasons, the evolution of the Montreal model was stopped at 89 Myrs, and the results for a 1.5M$_{\odot}$ model are the only ones presented. TGEC models with masses up to 2.5M$_{\odot}$ are computed along the whole main-sequence phase. Since the physics of our models and the r30-3M Montreal model are close but not similar, no detailed comparison between them can be carried out. However, qualitative similarities are underlined.

\subsection{Results}
\label{section:3.2}
\begin{figure}
  \resizebox{\hsize}{!}{\includegraphics{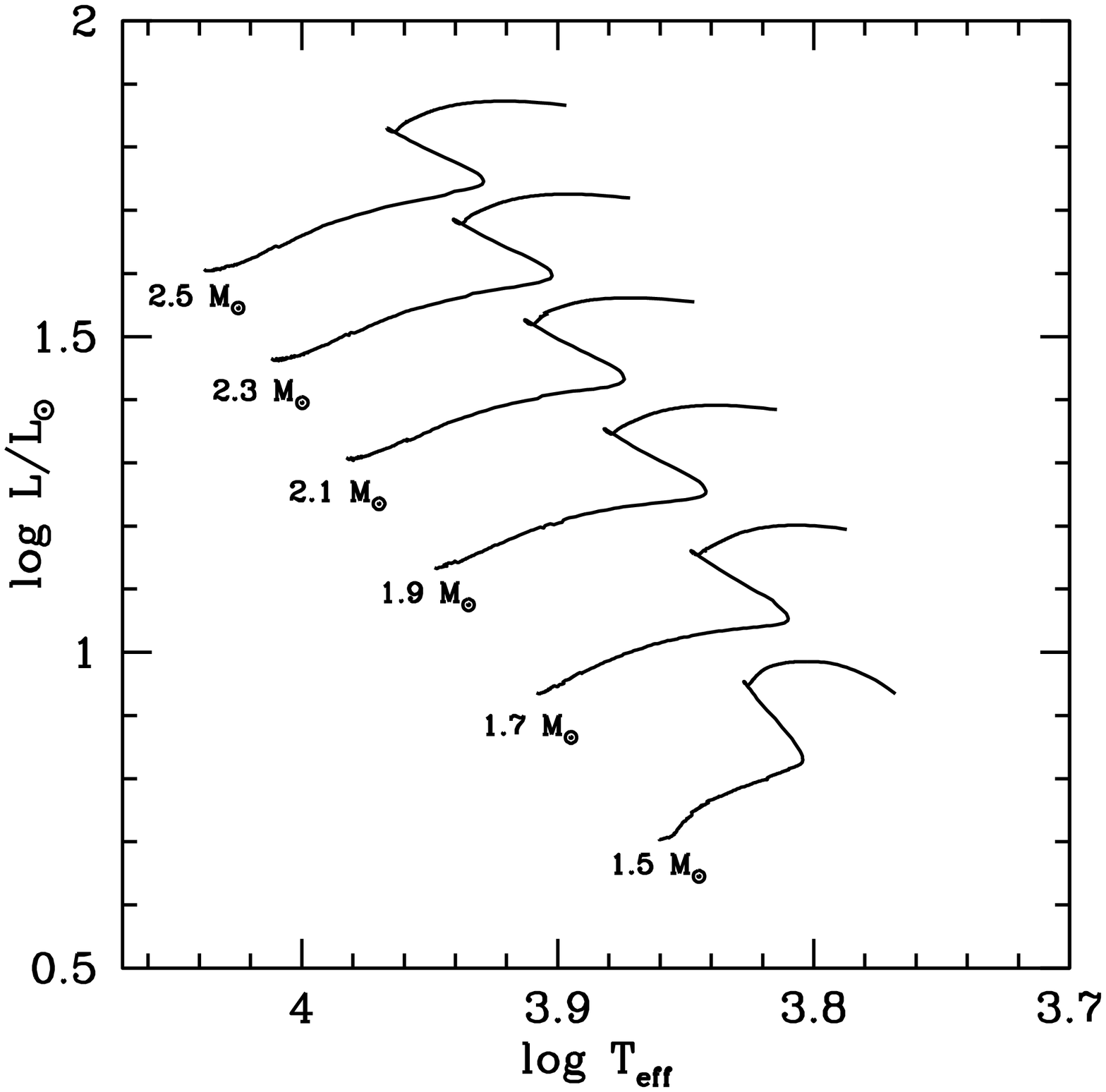}}
   \caption{Evolutionary tracks of 1.5 to 2.5M$_{\odot}$ models computed with atomic diffusion (including radiative acceleration effects) and a mild mixing below the convective enveloppe.}
  \label{diaghr}%
\end{figure}

Figure \ref{diaghr} displays the evolutionary tracks of our models. (For clarity  pre-main-sequence evolution is not represented.) In the following, we describe, as a representative example, the results obtained for a 2.1M$_{\odot}$ model. We focus on iron-diffusion related features.

\begin{figure*}
  \centering
  \includegraphics[width=0.32\textwidth]{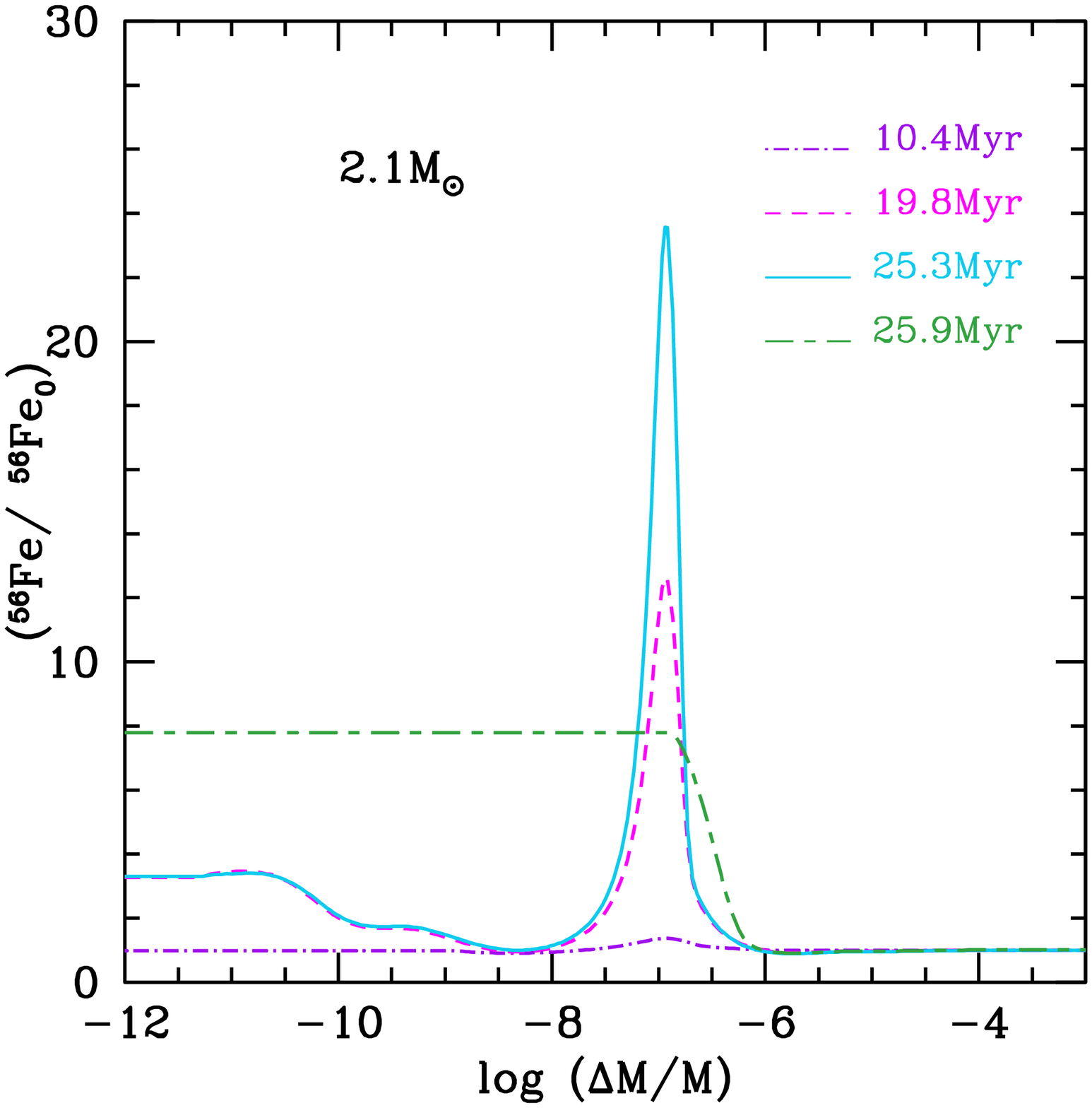}%
  \includegraphics[width=0.32\textwidth]{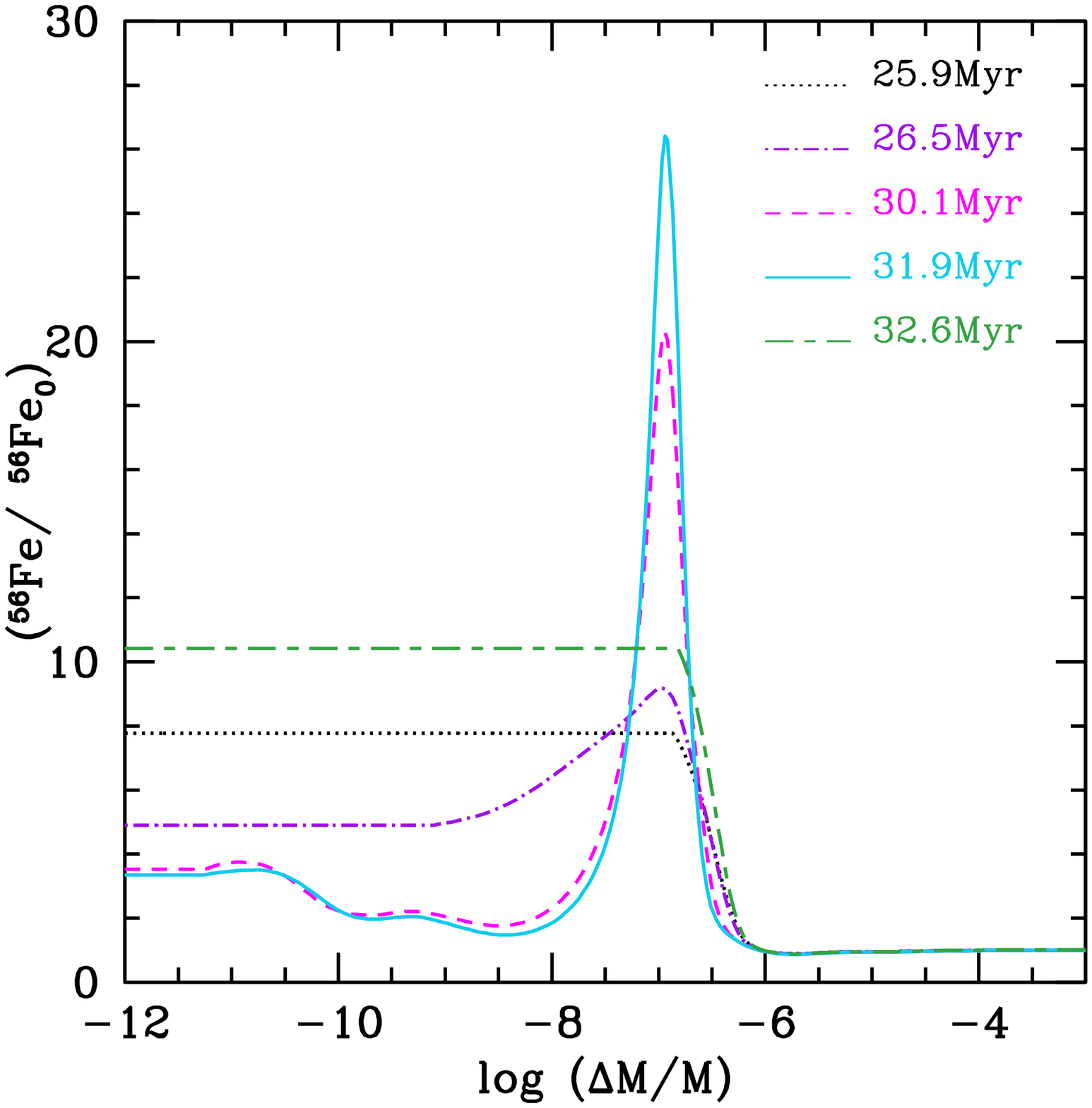}%
  \includegraphics[width=0.32\textwidth]{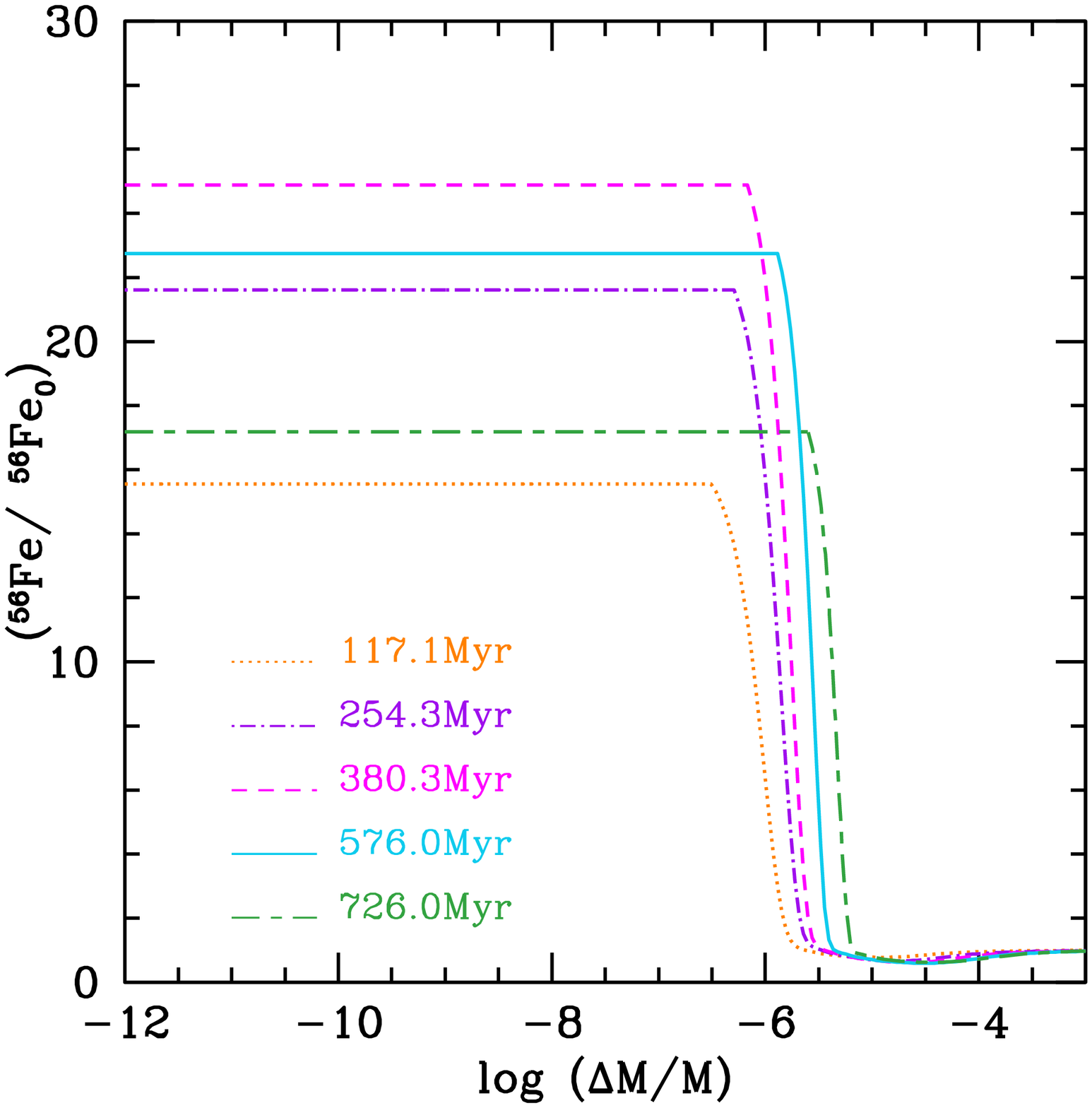}

  \includegraphics[width=0.32\textwidth]{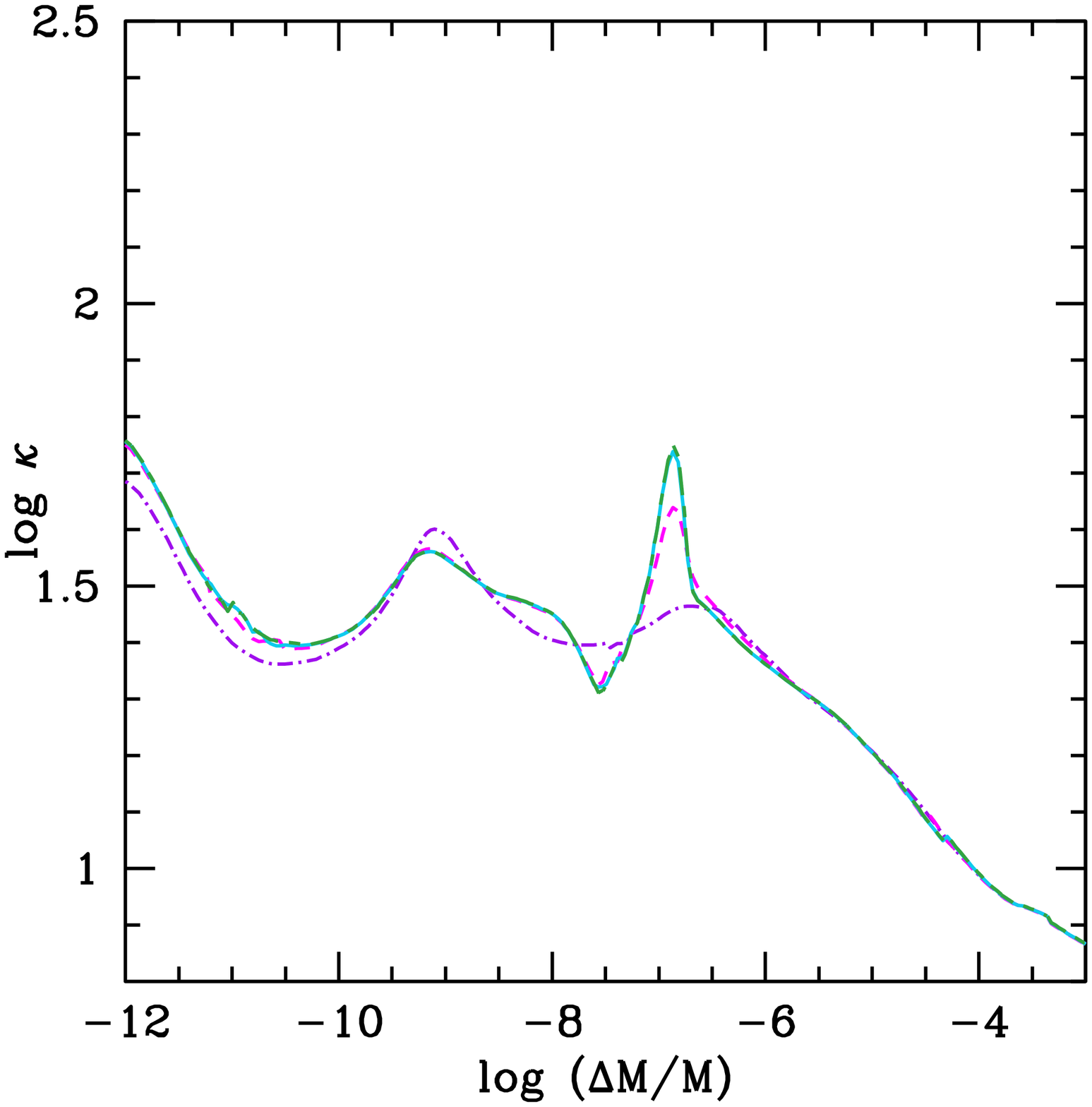}%
  \includegraphics[width=0.32\textwidth]{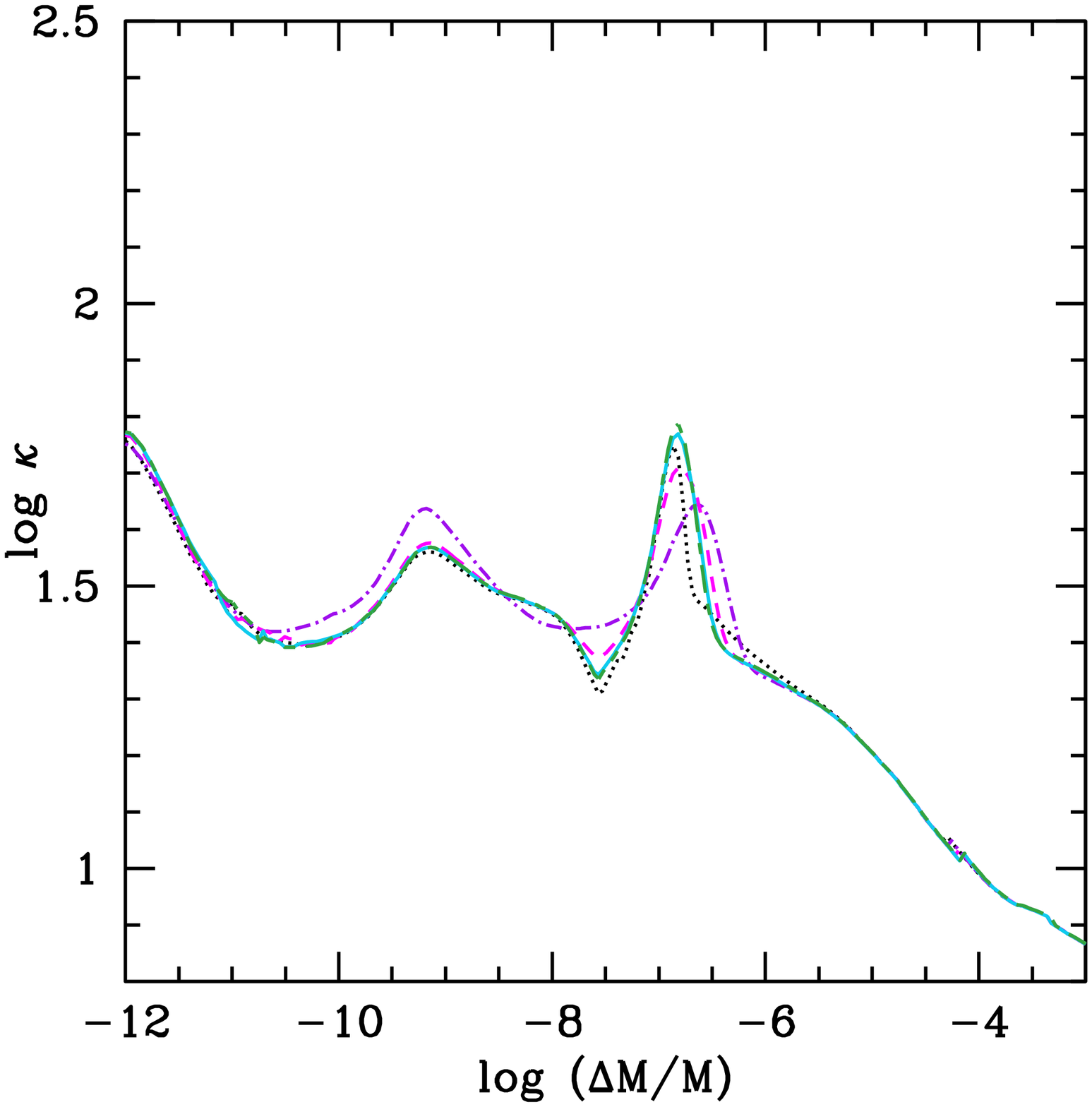}%
  \includegraphics[width=0.32\textwidth]{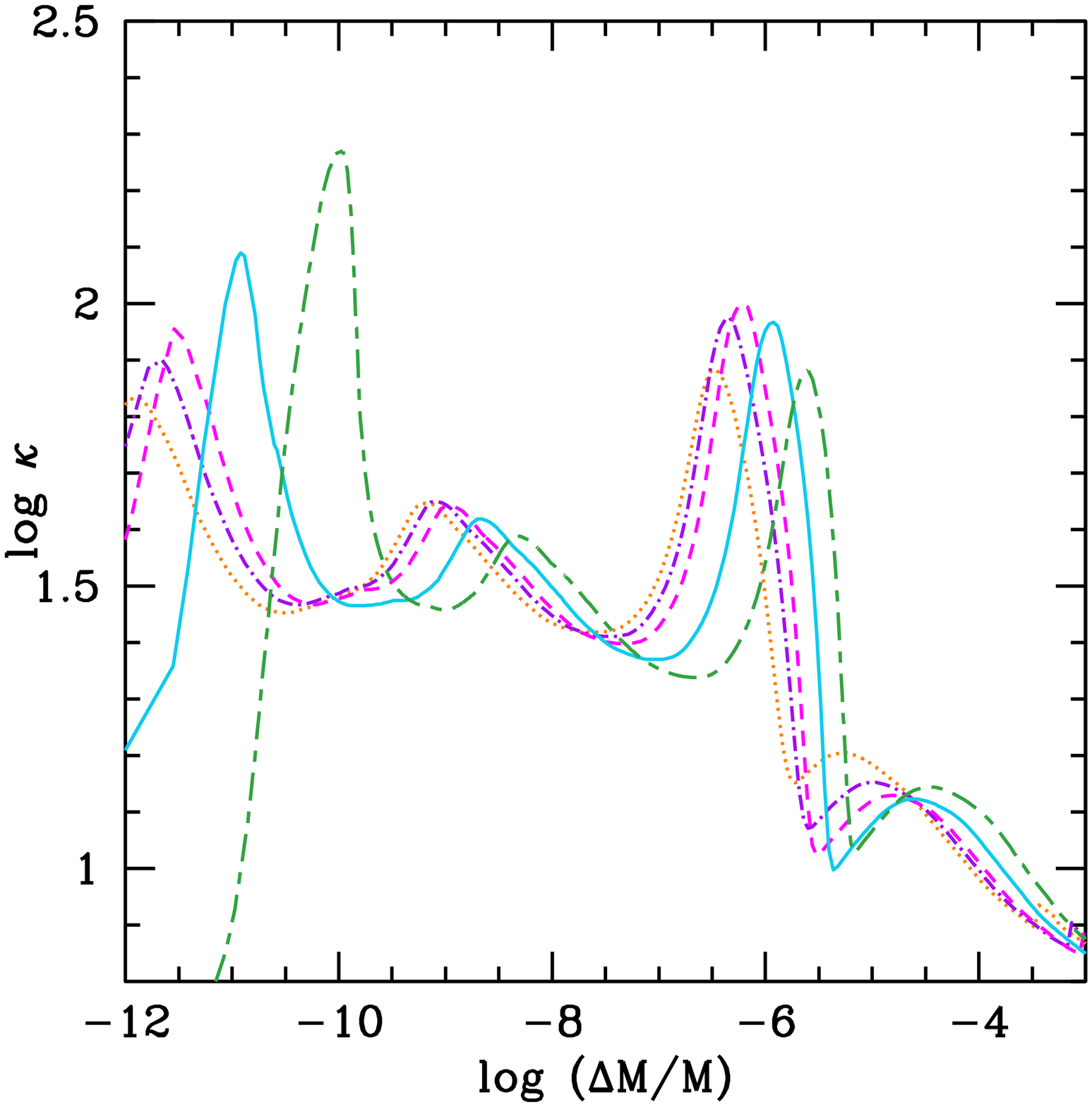}

  \includegraphics[width=0.32\textwidth]{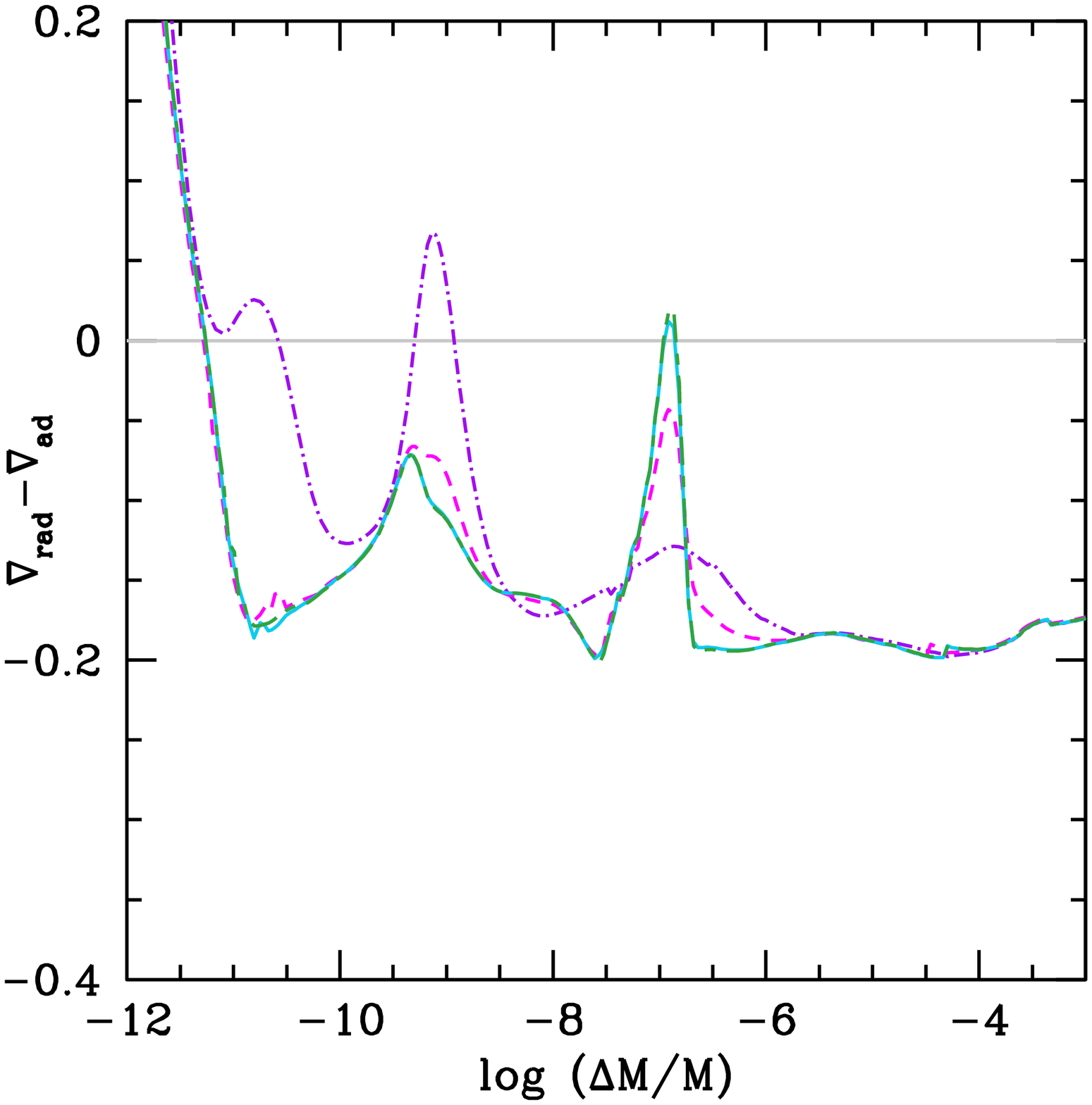}%
  \includegraphics[width=0.32\textwidth]{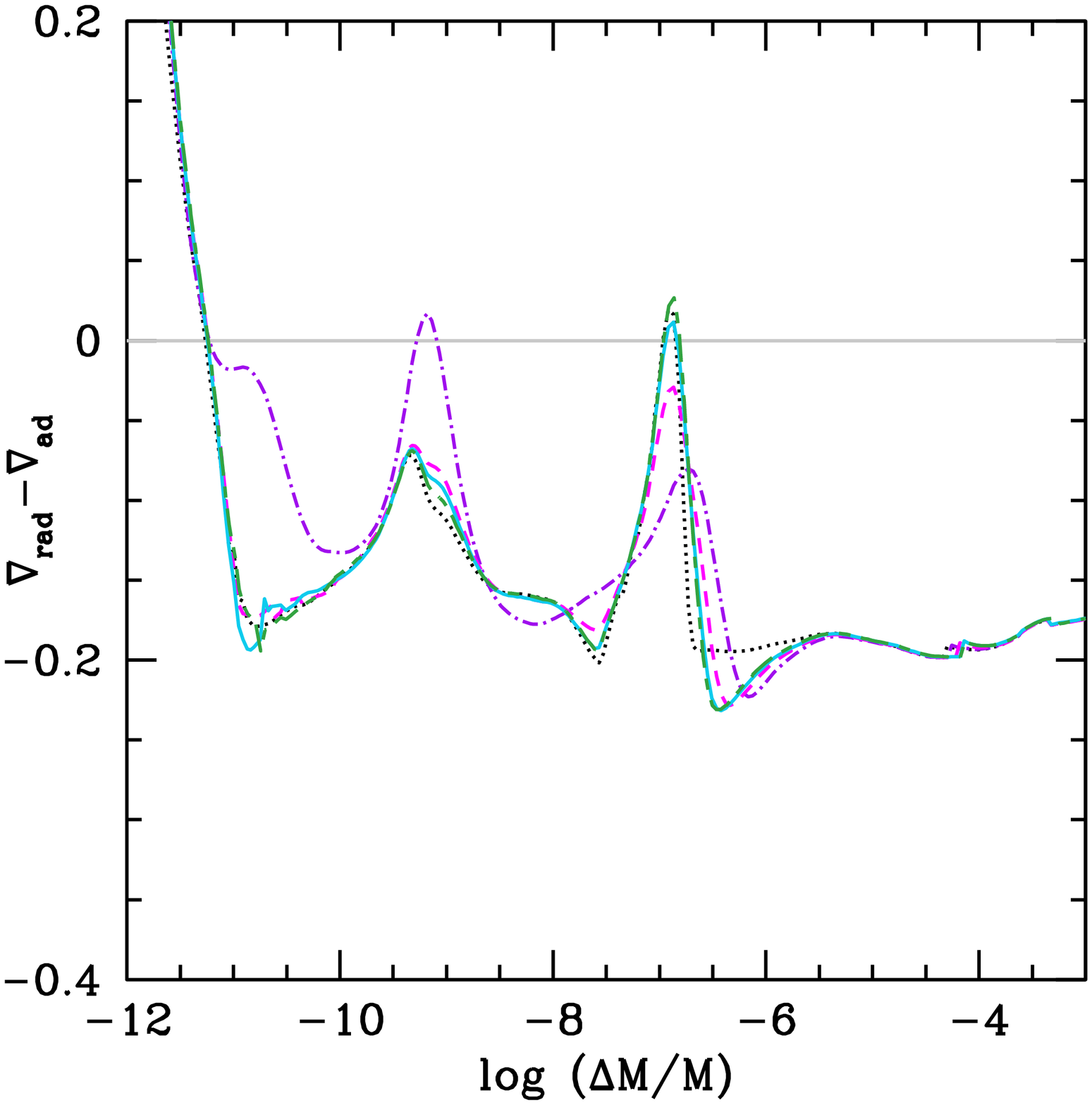}%
  \includegraphics[width=0.32\textwidth]{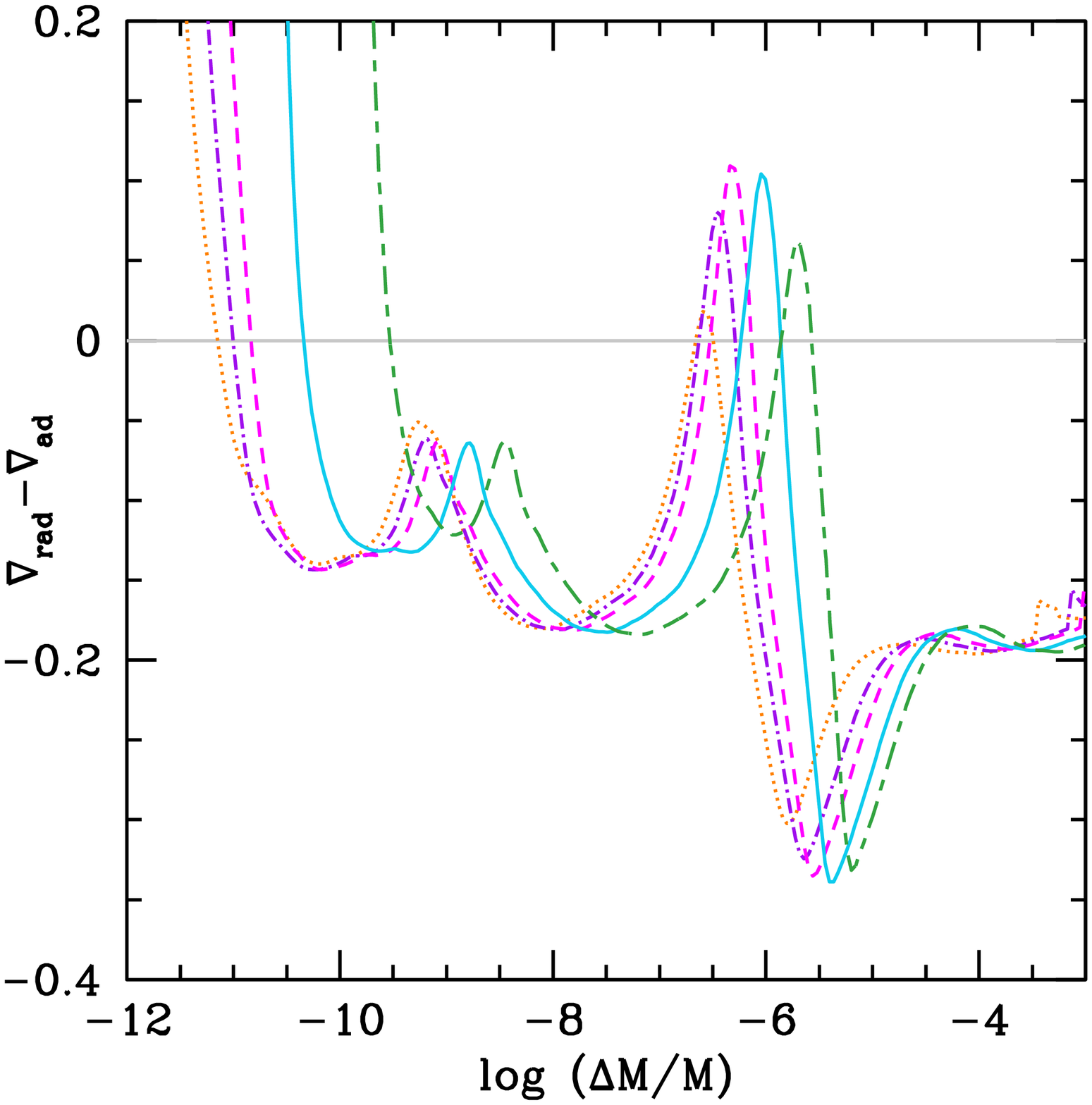}
  \caption{Internal structure of the 2.1M$_{\odot}$-model represented in Fig. \ref{diaghr}. The profiles are represented as a function of the outer mass fraction and at various evolutionary steps. Upper panel: iron abundance profiles, $\rm ^{56}Fe/^{56}Fe_0$ represents the ratio between the current Fe abundance and its initial value (cf. Table \ref{table1}). The various curves are defined in the upper panels and are valid for all the plots of the respective column. Middle panel: Rosseland opacity profiles. Lower panel: difference between the radiative and adiabatic gradients.}
  \label{masse21a}%
\end{figure*}

Figure \ref{masse21a} presents the profiles of various quantities inside the 2.1 M$_{\odot}$ model along the main-sequence evolution. The left and middle columns illustrate early main-sequence evolution, the right hand column displays later evolutionary stages.

On the main sequence, the competition between the gravitational settling and the radiative acceleration leads to iron accumulations in the outer regions of the model. We note iron enrichment in the surface H/He convective zone and in the Z-bump region (at around $\rm \log (\Delta M/M) =-7$ ($\rm T \simeq 200000K$). In this region, iron enrichment significantly increases the opacity, hence the radiative gradient, and when this radiative gradient exceeds the adiabatic gradient, the so-called iron convective zone appears. Here it is assumed to be connected and mixed up to the H/He surface convective zone through overshooting. As a result, the bottom of the surface convective zone sinks abruptly to $\rm \log (\Delta M/M) \simeq -7$. The dilution in this thick convective zone decreases the iron abundance in the opacity bump: the opacity and the radiative gradient consequently decrease, which leads to the disappearance of the iron convective zone. The surface convective zone then recedes. A new cycle starts as radiative forces proceed in accumulating iron in the opacity bump region. The middle column of Fig. \ref{masse21a} shows a second iron accumulation/deep convection phase.

The alternation between convective and radiative episodes in the Z-bump proceeds from $\simeq$20 to $\simeq$115 Myr. A thick convective zone then stays for the subsequent main-sequence evolution. The right hand column of Fig. \ref{masse21a} illustrates the internal structure variations of our model during the later evolutionary phases.

\begin{figure*}
  \centering
  \includegraphics[width=0.4\textwidth]{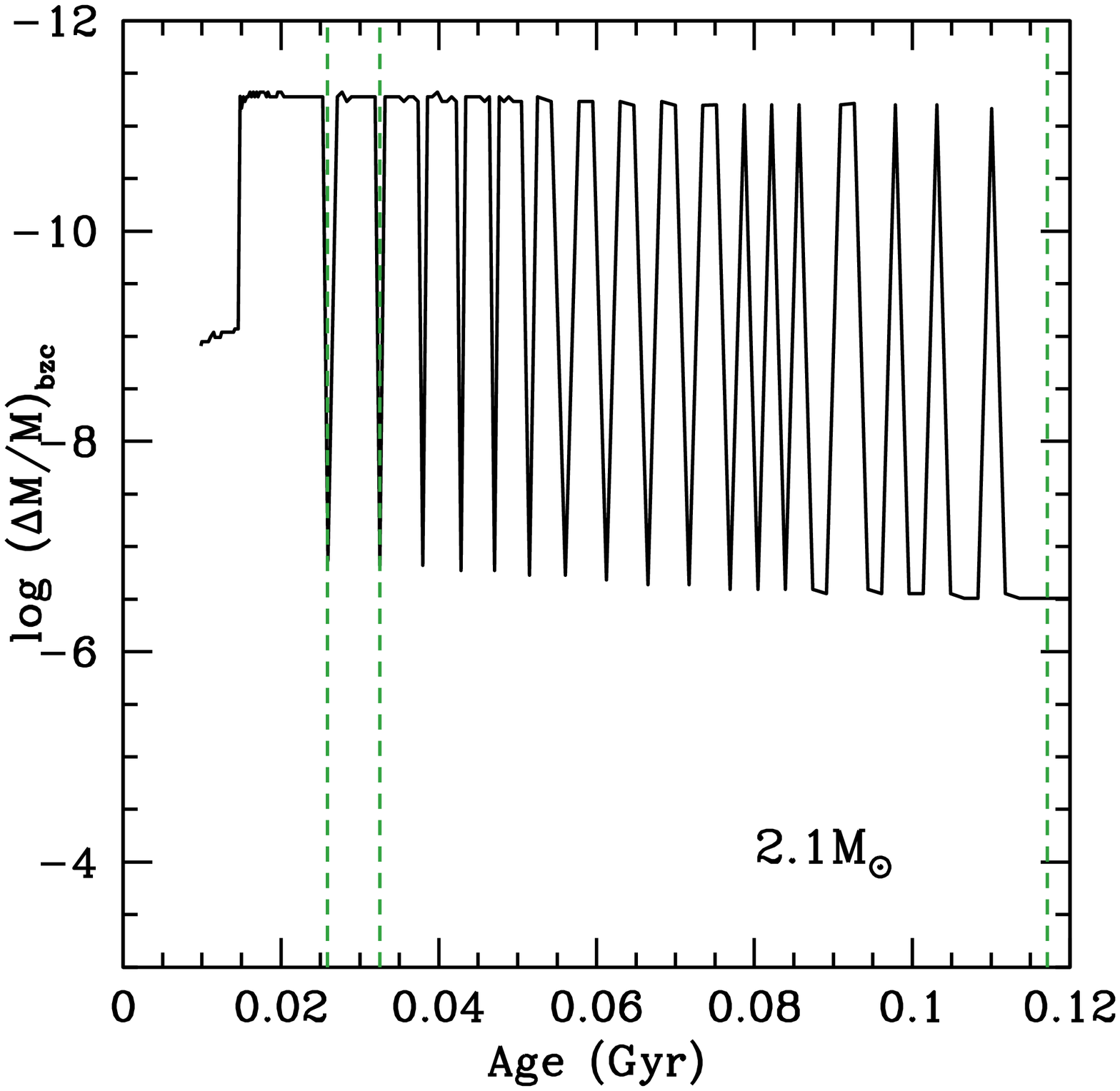}%
  \includegraphics[width=0.4\textwidth]{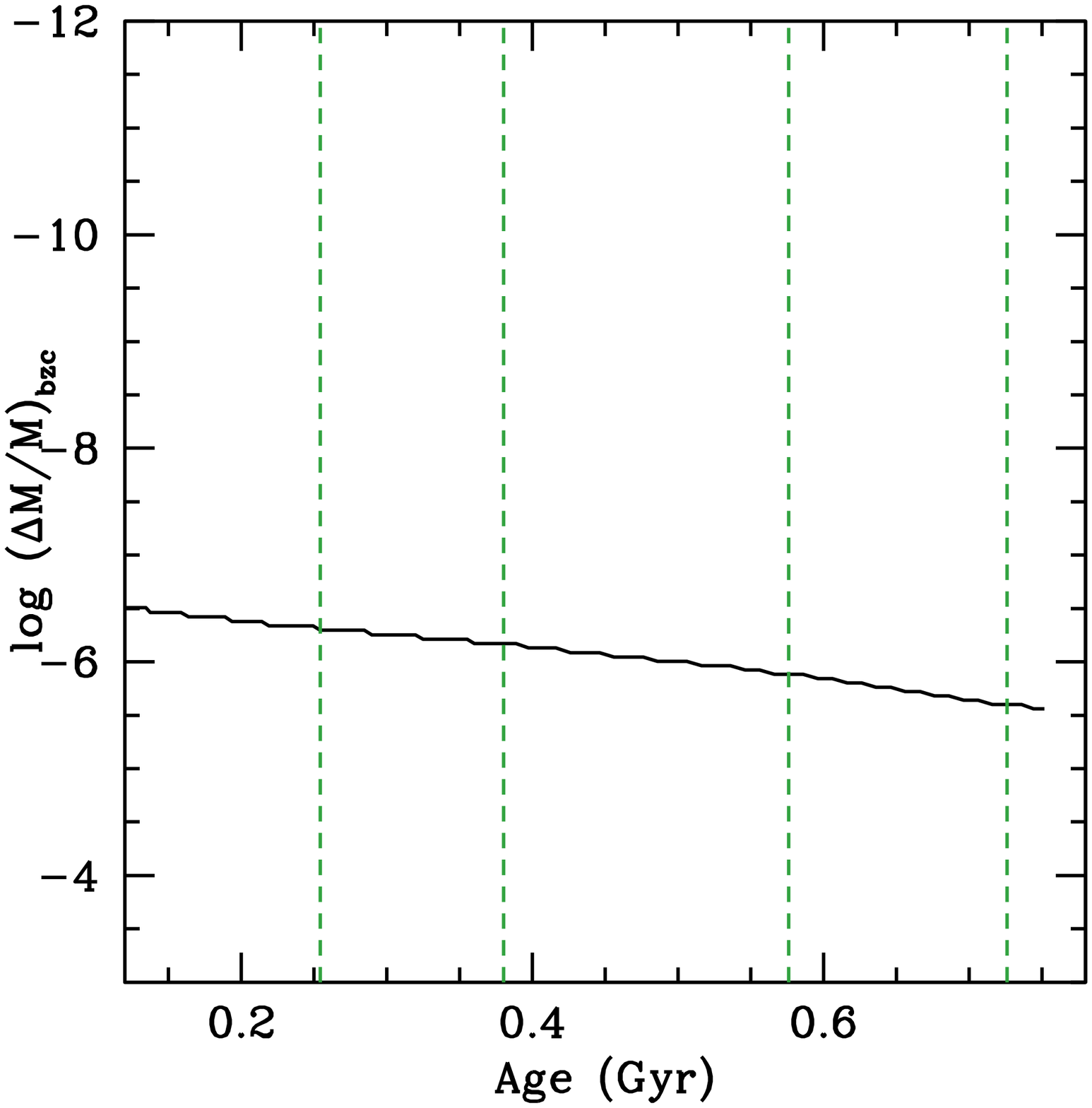}%
  \caption{Variation with time of the bottom of the surface convective zone in the 2.1M$_{\odot}$ model represented in Fig. \ref{diaghr}. Left panel: zoom on the 0-0.12Gyr period. Right panel: zoom on the 0.12-0.8Gyr period. The dashed vertical lines are located respectively at 25.9Myr, 32.6Myr, 117.1Myr, 254.3Myr, 380.3Myr, 576.0Myr, and 726.0Myr.}
  \label{zconvmzoom}%
\end{figure*}

Figure \ref{zconvmzoom} shows the position of the bottom of the surface convective zone of our model during the main sequence. At the beginning of the main-sequence phase, the convective zone includes the H and He ionization regions. After a few million years and because of the He gravitational settling, the HeII convective zone disappears leading to an abrupt recession of the convective region (whose bottom subsequently lies at $\rm \log (\Delta M/M) \simeq-11.25$). After this first receding episode, the convective region undergoes a 100Myr-period of rapid variations during which the convective depth oscillates between $\rm \log (\Delta M/M) \simeq -11.2$ and $\rm \log (\Delta M/M) \simeq -6.5$. After this period, a thick surface convective zone appears that slowly deepens inside the interior during the rest of the main-sequence phase.  

The lasting thick convective zone results from two features explained below.
\begin{itemize}
\item Because of structural variations forced by the star's evolution, the iron accumulation needed to induce convection in the opacity bump varies along the main sequence. After 115 Myrs, in particular, smaller iron accumulations in the Z-bump are needed to initiate the iron convective zone. 

\item The succession of convective sinking and receding episodes, which occurs from 20 to $\simeq$115Myr, leads to rapid variations in the surface iron abundance. As an illustration of this phenomenom, Fig. \ref{fesurf50} displays the iron abundance evolution over 30Myrs (including 5 convective sinking/receding episodes).
\begin{figure}
\centering
\includegraphics[width=0.5\textwidth]{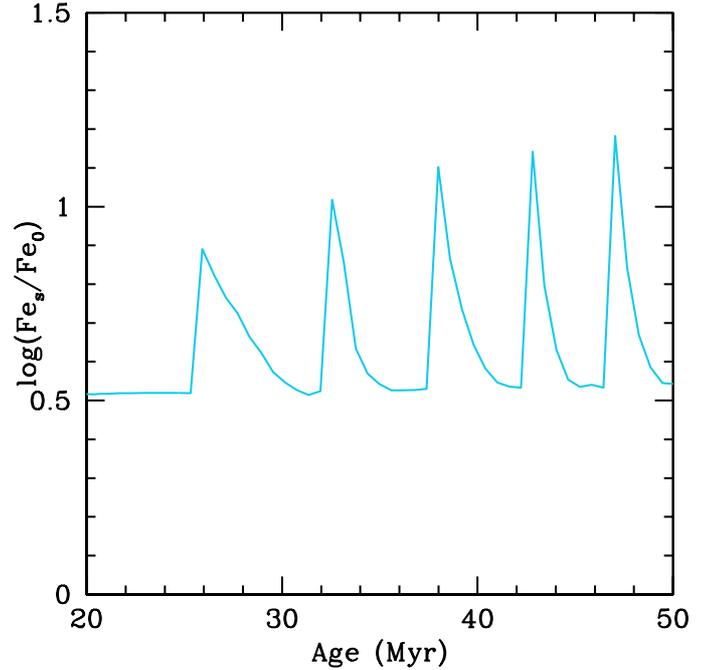}
\caption{Time dependent variations in the Fe surface abundance in the 2.1M$_{\odot}$ model presented in Fig. \ref{diaghr}. The iron abundance is shown on a 30Myr period including 5 convective sinking/receding episodes. This graph is similar to Fig. 16 of \citet{Richard01} (see text for more details).}
\label{fesurf50}
\end{figure}
The global effect of these episodes is illustrated in Fig. \ref{fesurf}, which presents the time-dependent variations of the averaged iron surface abundance (averaged over 20 Myrs).  
\begin{figure}
  \resizebox{\hsize}{!}{\includegraphics{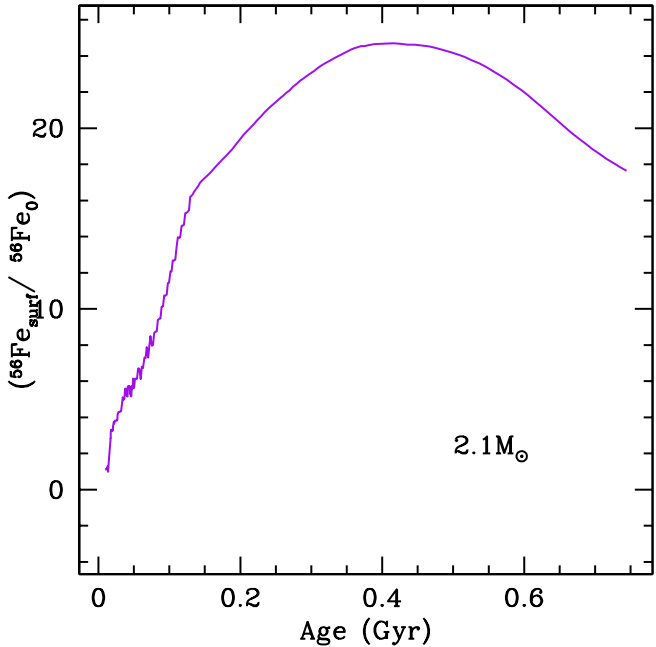}}
   \caption{Iron surface abundance along main-sequence evolution for the 2.1M$_{\odot}$ model represented in Fig. \ref{diaghr}. During the 0-120Myr period (i.e. when the external convective zone undergoes rapid and large variations), the iron abundance shown is an averaged value (over 20Myr).}
  \label{fesurf}%
\end{figure}
The diffusion-induced iron enrichments (in the Z-bump but also in the H/He convective zone) combined with the deep convective mixing episodes leads on average to an iron enrichment of the surface convective zone.
 For high values of this enrichment, the convective dilution in the presence of an iron convective zone may become unable to efficiently decrease the iron abundance in the Z-bump and to suppress the iron convective zone. A thick convective zone then persists during the subsequent evolution. 
\end{itemize}

According to Fig. \ref{fesurf}, the iron abundance in the surface convective zone continues to increase after 115 Myrs (because of radiative accelerations), while the convective zone slowly deepens. After 400Myr, the bottom of the convective zone reaches iron-underabundant regions, which causes as a slow decrease in the iron surface abundance. 

\begin{figure*}
  \centering
  \includegraphics[width=0.32\textwidth]{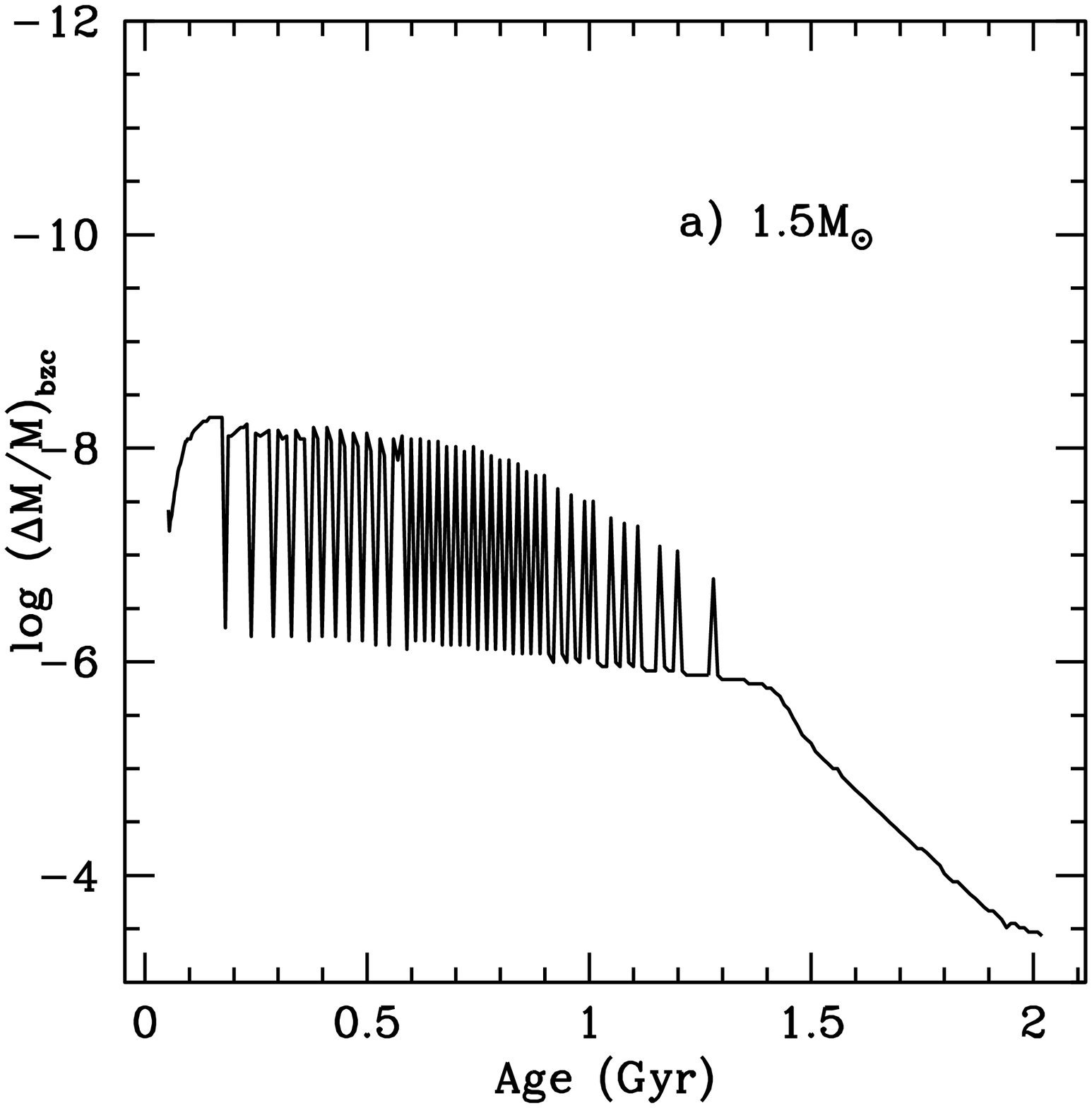}%
  \includegraphics[width=0.32\textwidth]{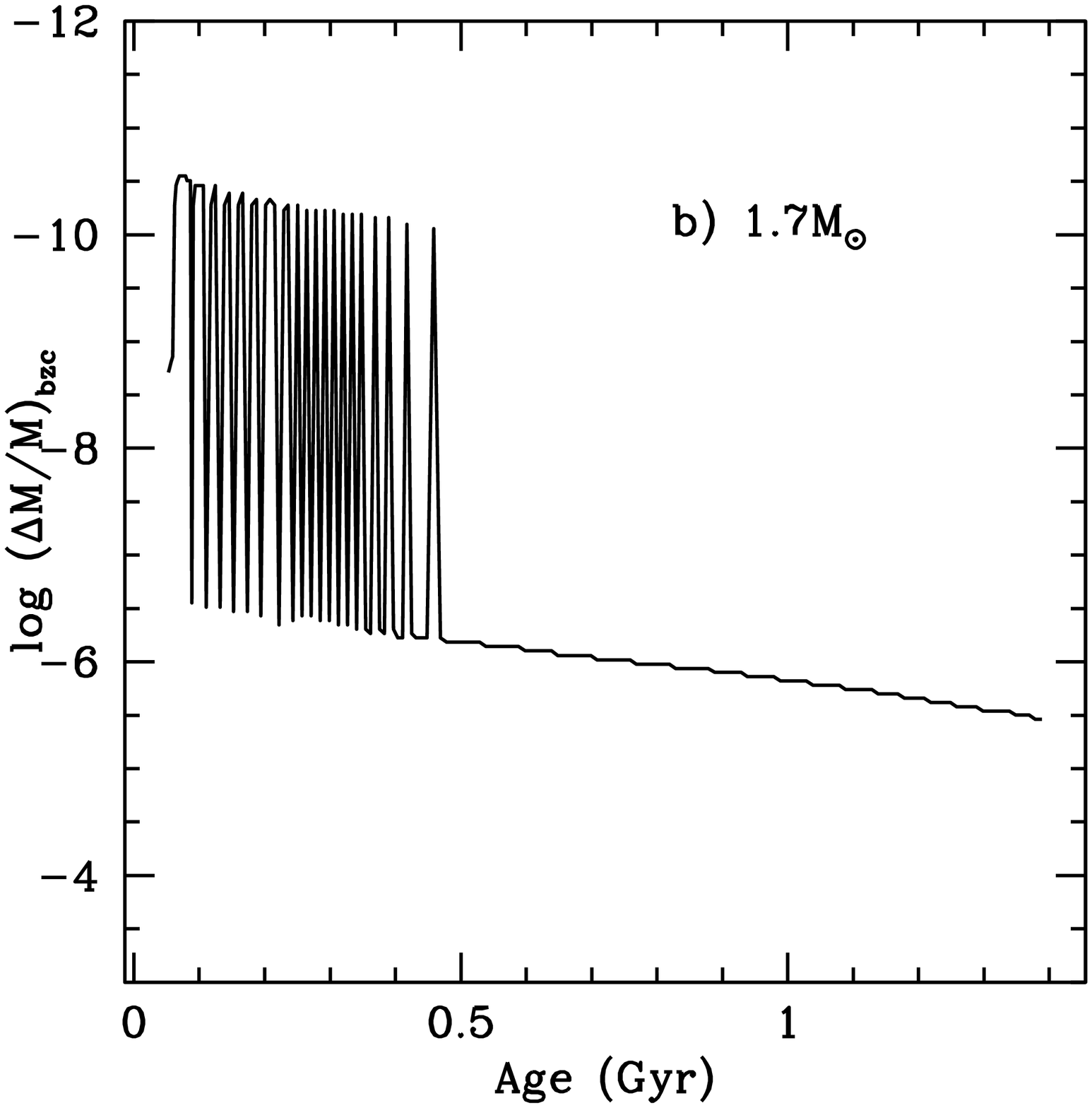}%
  \includegraphics[width=0.32\textwidth]{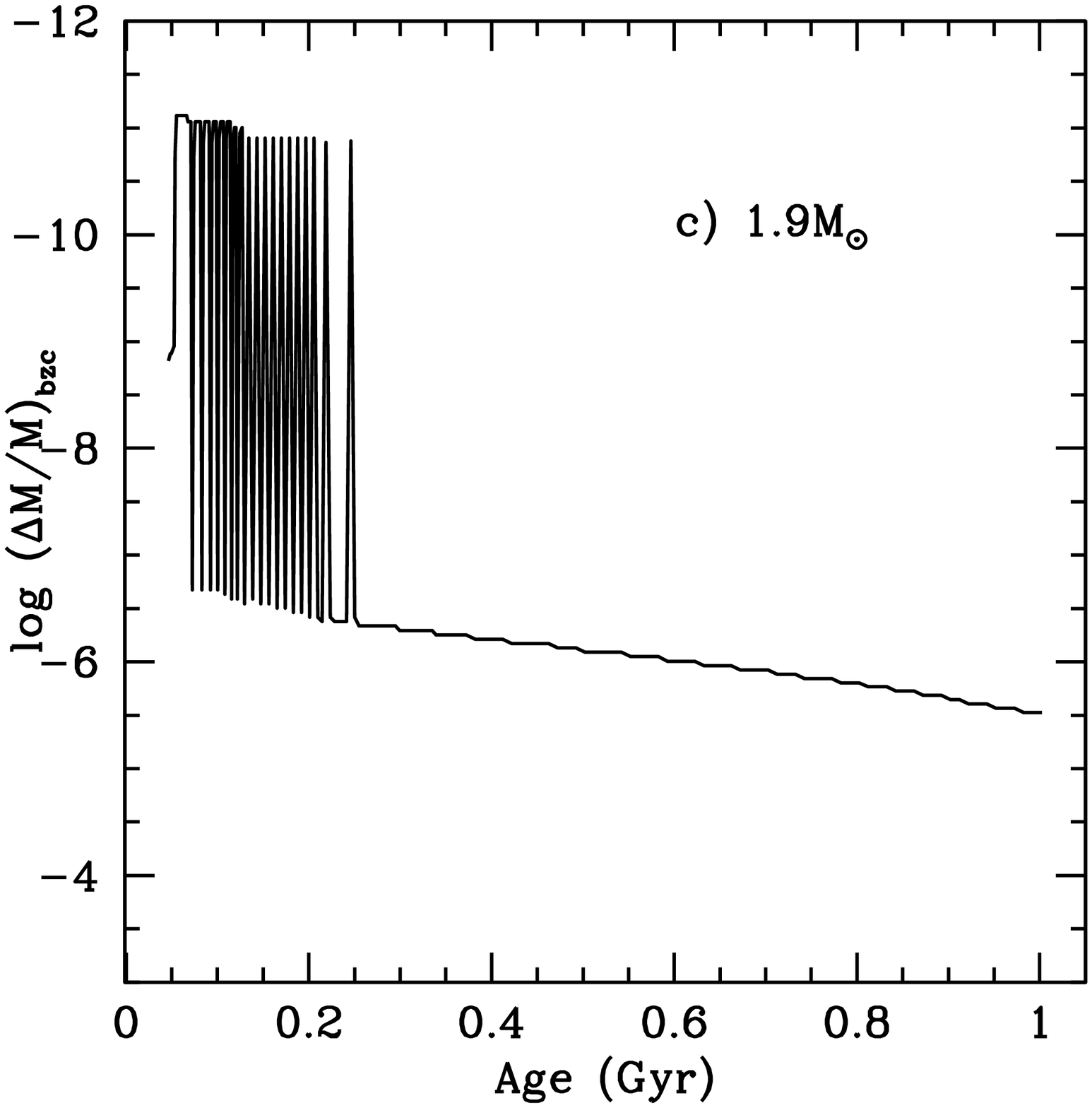}
  \includegraphics[width=0.32\textwidth]{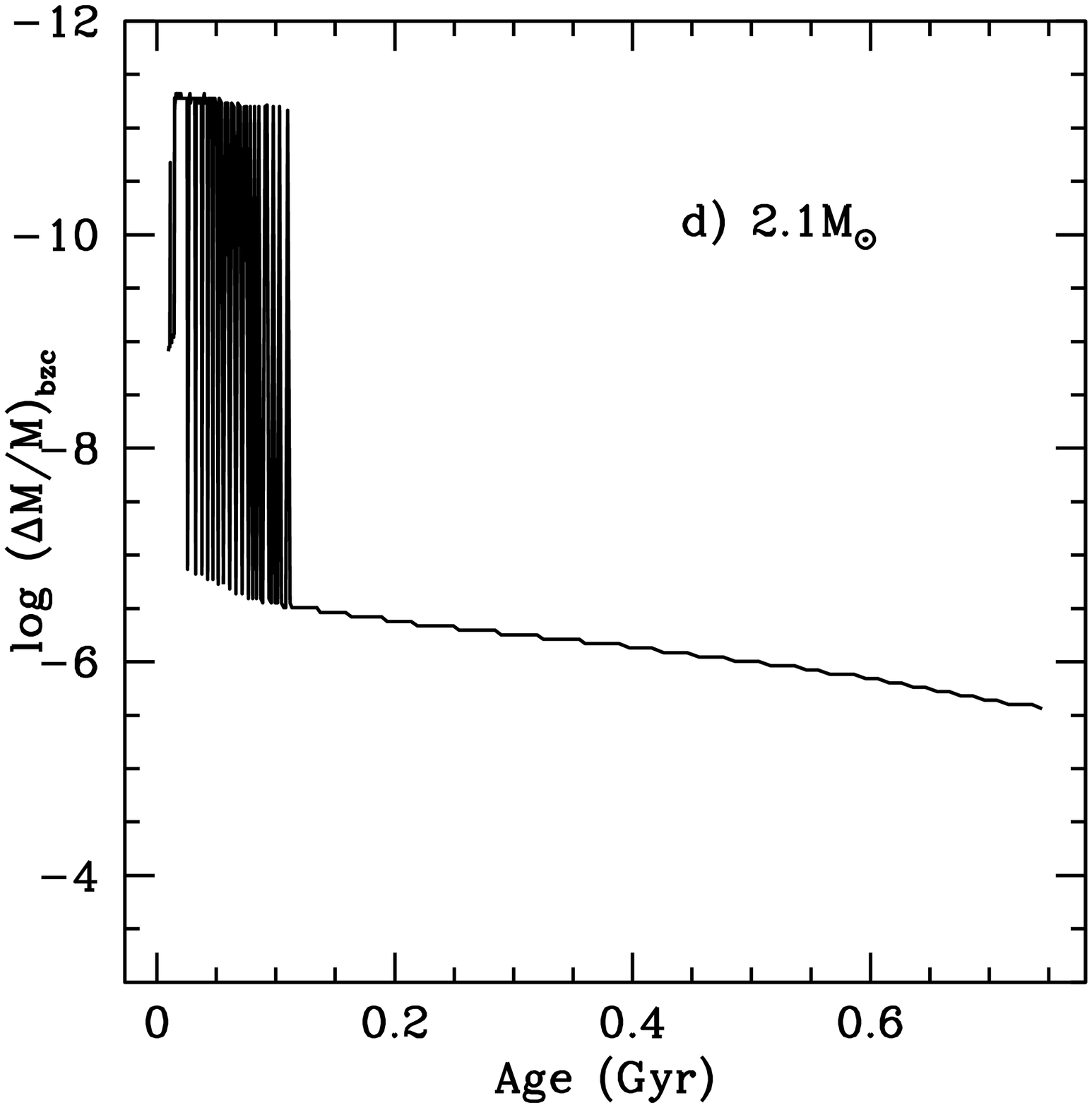}%
  \includegraphics[width=0.32\textwidth]{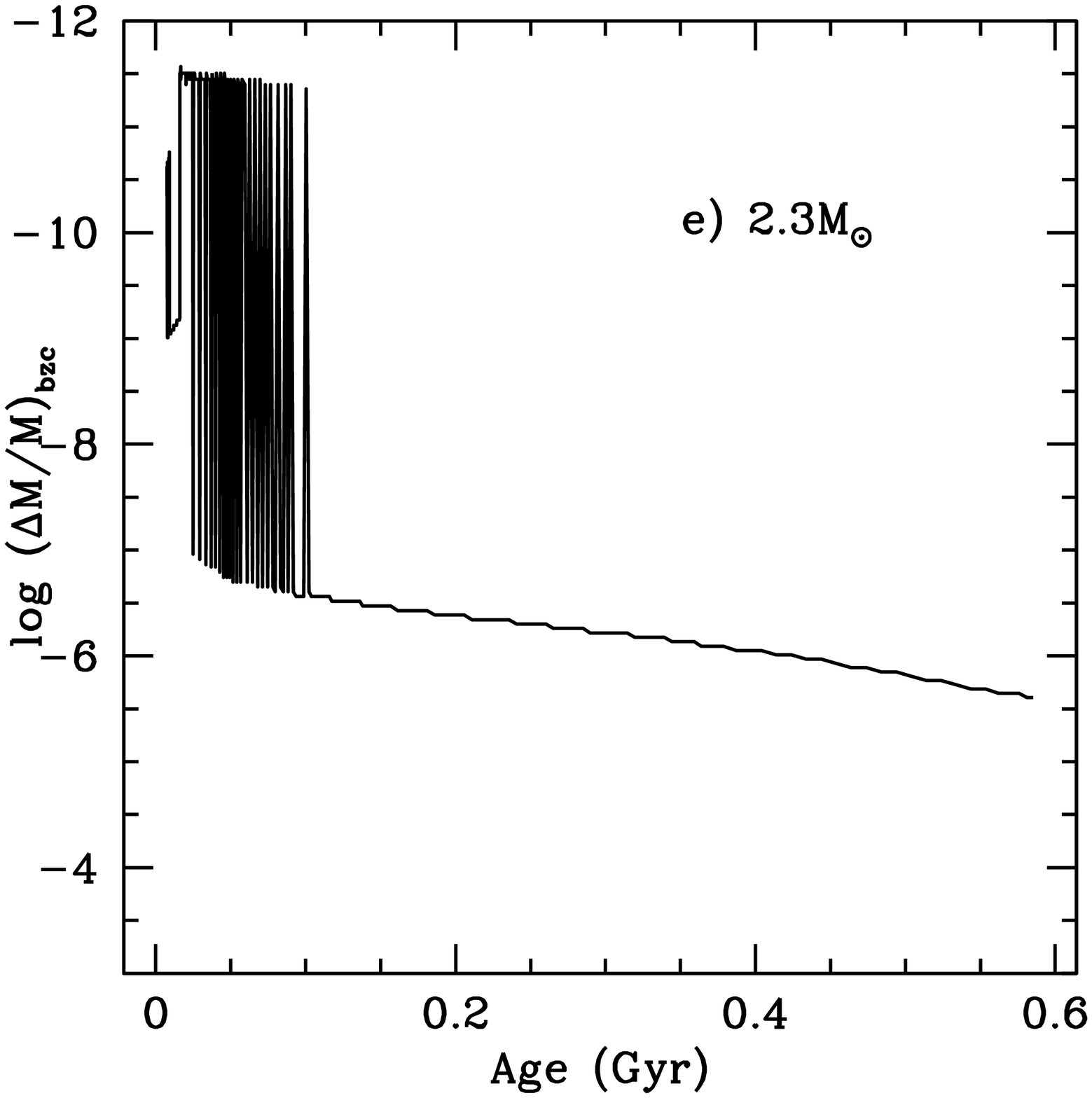}%
  \includegraphics[width=0.32\textwidth]{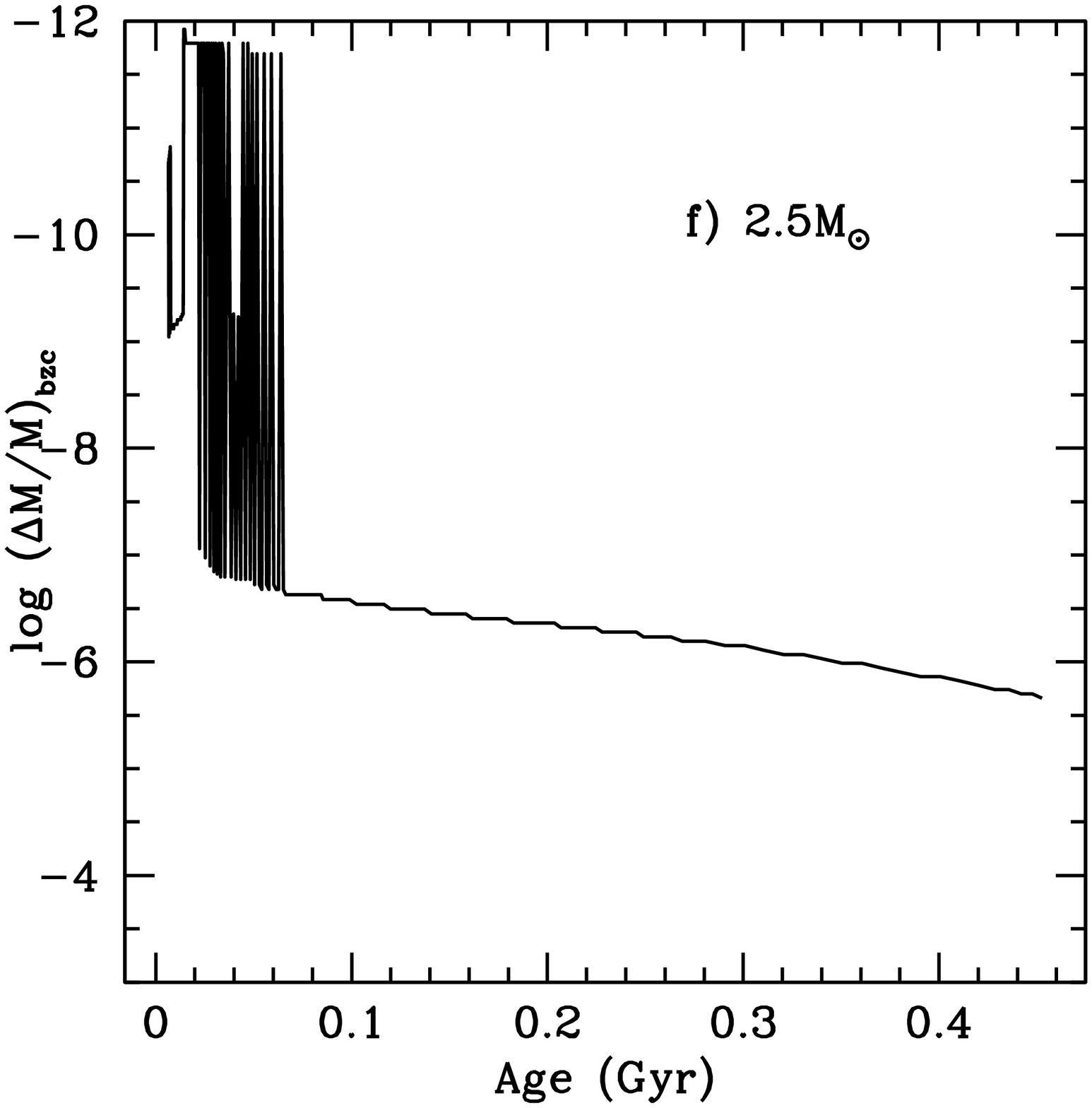}
  \caption{Variation in the surface convective zone extension along the main-sequence phase for the same models as presented in Fig. \ref{diaghr}.  The vertical axis represents the outer mass fraction, the solid curves represent the bottom of the external convective zone.}
  \label{zconvm}%
\end{figure*}

Figure \ref{zconvm} presents the convective zone extension (vs time) in all computed models. For an easy comparison with other masses, the 2.1 M$_{\odot}$ model is shown in this figure. A succession of convective sinking and receding episodes is observed in all the computed models. This period of rapid oscillations of the convective depth is shorter for higher stellar masses. In the 1.5 M$_{\odot}$ model, it persists during most of the main-sequence phase, whereas in the 2.5 M$_{\odot}$ model, a thick convective zone appears in less than 80 Myrs.

The patterns observed during the early main-sequence phase of our models are consistent with the results presented by \cite{Richard01} for their r30-3M, 1.5M$_{\odot}$ model. The alternation between radiative and convective episodes in the opacity bump that occurs in our models is similarly observed in the Montreal model as described in Sect. 3 (paragraph ``Model with mixing episodes'') of \cite{Richard01}. The comparison between our Fig. \ref{fesurf50} and Fig. 16 of \cite{Richard01} (middle panel) also shows similar behaviour to the surface iron abundances. But while the Montreal code is limited by time consuming issues (their r30-3M, 1.5M$_{\odot}$ model is evolved for less than 100 Myr), TGEC is able to manage the rapid structural and compositional variations due to the succession of convective sinking and receding episodes and allows computing complete main-sequence evolutionary tracks for masses up to at least 2.5M$_{\odot}$.

\section{Conclusion}
The implementation of radiative diffusion effects in stellar modeling turns out as a necessary condition for any computations of accurate models of F, A, and B type stars. Different ways of introducing these effects are considered by theoreticians, but few reliable methods are presently known.  

The most accurate method consists in computing the time-dependent variations of the chemical species present in the stellar mixture while using a precise description of the diffusion process. This includes an accurate determination of the radiative accelerations applied to each species and the resolution of the Burgers equations to obtain the diffusion velocities. Such a method has been successfully introduced in the Montreal code, which has been used for many years for the modeling of main-sequence F-A-B type stars. This accurate method allows precise determinations of abundance variations due to atomic diffusion, but it leads to highly time-consuming codes, which makes the computations of complete main-sequence evolutionary tracks difficult. 

The development of asteroseismology during the past 30 years and, in particular, the discovery of a wide variety of pulsating F, A, and B-type stars have brought to light the necessity for developing stellar modeling tools able to provide accurate models adapted to seimic analysis for these stars. The seismic modeling of a target star is an iterative process that requires computating numerous models before getting the best one. In this context, using a code like the Montreal one appears particularly unadapted.

This encouraged us to develop an alternative code that includes a more flexible prescription of the radiative accelerations. In this context we included the radiative diffusion in the TGEC code using the semi-analytical prescription proposed by \citet{Alecian02} and \citet{LeBlanc04} (called the SVP method). This method, less accurate than the one introduced in the Montreal code, has been proven to give good results with reasonable precision. With the most recent version of TGEC, this method has been successfully introduced for the first time in a stellar evolution code. In TGEC, the diffusion velocities are then computed following the Chapman \& Cowling formalism in the test-atom approximation. The time-dependent variations in the chemical species are computed consistently. 

To validate our code, we did a comparison between a model computed with the Montreal code and a similar model computed with TGEC. (The two models include input physics that are as close as possible.) The comparison shows good agreement in the computations of the radiative acceleration for models with similar internal structures.

Nickel is not presently included in the TGEC computations, as its atomic data are not included in TOPBase \citep{Cunto93}, the atomic data base of the Opacity Project \citep{Seaton92} from which the atomic data used in the SVP approximation are taken. This explains its absence in the SVP package. Its introduction in TOPBase\footnote{Nickel is, however, included in the opacity calculation of Opacity Project in an approximate way.} as well as in the SVP package, is underway.

To demonstrate both the abilities of the TGEC code and its flexibility as compared to the Montreal one, we presented sets of models that include atomic diffusion with small additional mixing. In this context, models including ``minimal mixing'' were already presented in \citet{Theado09}; in Sect. \ref{computations1} we chose to present a new set of models (with masses ranging from 1.5 to 2.5M$_{\odot}$) including physical ingredients close to the r30-3M, 1.5M$_{\odot}$ model of \citet{Richard01}. In these models, only mild mixing is introduced below the surface convective zone, and the iron convection region, when its appears, is assumed to be connected and mixed with the surface convective envelope. Under these assumptions, our models show similar behavior to the Montreal models, at least during the first 100Myr (where the Montreal computations stop). In particular, the introduced physical hypothesis leads to an alternation of convective and radiative episodes in the opacity bump region in both the Montreal and the TGEC models. We have also shown that the TGEC code allows computing complete evolutionary tracks (up to the subgiant branch) for A-F stars with masses up to 2.5M$_{\odot}$.
It is able to manage rapid variations in the chemical composition, under cover of mild mixing at the transition between convective and radiative regions. The effects of quasi-pure atomic diffusion can then be evaluated in the context of stellar evolutionary models. 

As discussed in the introduction, precise tests of the results obtained by the two codes for the case of iron accumulation inside stars, when thermohaline convection is not taken into account, still have to be performed, since differences in the maximum iron value appear in some cases. Such tests are underway and will be presented elsewhere in the near future.

The improvements brought to the atomic diffusion aspects of the TGEC code presented here and in \citet{Theado09} makes it an excellent test bed for stellar evolution and asteroseismology. Its application to various types of stars will certainly help us understand them better.

\begin{acknowledgements}
We thank Olivier Richard for kindly providing its 5.3D50-3 1.70 M$_{\odot}$ model. We acknowledge the financial support of the Programme National de Physique Stellaire (PNPS) of CNRS/INSU, France and the Natural Sciences and Engineering Research Council of Canada.
\end{acknowledgements}


\begin{thebibliography}{38}
\expandafter\ifx\csname natexlab\endcsname\relax\def\natexlab#1{#1}\fi

\bibitem[{{Alecian}(1985)}]{AlecianAl1985}
{Alecian}, G. 1985, A\&A, 145, 275

\bibitem[{{Alecian}(1994)}]{AlecianAl1994}
{Alecian}, G. 1994, A\&A, 289, 885

\bibitem[{{Alecian} \& {Artru}(1990)}]{AlecianAlAr1990a}
{Alecian}, G. \& {Artru}, M. 1990, A\&A, 234, 323

\bibitem[{{Alecian} \& {LeBlanc}(2002)}]{Alecian02}
{Alecian}, G. \& {LeBlanc}, F. 2002, \mnras, 332, 891

\bibitem[{{Angulo}(1999)}]{Angulo99}
{Angulo}, C. 1999, in American Institute of Physics Conference Series, Vol.
  495, American Institute of Physics Conference Series, 365--366

\bibitem[{{Asplund} {et~al.}(2005){Asplund}, {Grevesse}, \&
  {Sauval}}]{Asplund05}
{Asplund}, M., {Grevesse}, N., \& {Sauval}, A.~J. 2005, in Astronomical Society
  of the Pacific Conference Series, Vol. 336, Cosmic Abundances as Records of
  Stellar Evolution and Nucleosynthesis, ed. {T.~G.~Barnes III \& F.~N.~Bash},
  25

\bibitem[{{Asplund} {et~al.}(2009){Asplund}, {Grevesse}, {Sauval}, \&
  {Scott}}]{Asplund09}
{Asplund}, M., {Grevesse}, N., {Sauval}, A.~J., \& {Scott}, P. 2009, \araa, 47,
  481

\bibitem[{{Bahcall} \& {Pinsonneault}(1992)}]{Bahcall92}
{Bahcall}, J.~N. \& {Pinsonneault}, M.~H. 1992, Reviews of Modern Physics, 64,
  885

\bibitem[{{Brun} {et~al.}(1998){Brun}, {Turck-Chi{\`e}ze}, \& {Morel}}]{Brun98}
{Brun}, A.~S., {Turck-Chi{\`e}ze}, S., \& {Morel}, P. 1998, \apj, 506, 913

\bibitem[{{Burgers}(1969)}]{Burgers69}
{Burgers}, J.~M. 1969, {Flow Equations for Composite Gases}, ed. {Burgers,
  J.~M.}

\bibitem[{{Chapman} \& {Cowling}(1970)}]{Chapman70}
{Chapman}, S. \& {Cowling}, T.~G. 1970, {The mathematical theory of non-uniform
  gases. an account of the kinetic theory of viscosity, thermal conduction and
  diffusion in gases}, ed. {Chapman, S.~\& Cowling, T.~G.}

\bibitem[{{Christensen-Dalsgaard} \& {Daeppen}(1992)}]{JCD92}
{Christensen-Dalsgaard}, J. \& {Daeppen}, W. 1992, \aapr, 4, 267

\bibitem[{{Christensen-Dalsgaard} {et~al.}(1996){Christensen-Dalsgaard},
  {Dappen}, {Ajukov}, {Anderson}, {Antia}, {Basu}, {Baturin}, {Berthomieu},
  {Chaboyer}, {Chitre}, {Cox}, {Demarque}, {Donatowicz}, {Dziembowski},
  {Gabriel}, {Gough}, {Guenther}, {Guzik}, {Harvey}, {Hill}, {Houdek},
  {Iglesias}, {Kosovichev}, {Leibacher}, {Morel}, {Proffitt}, {Provost},
  {Reiter}, {Rhodes}, {Rogers}, {Roxburgh}, {Thompson}, \& {Ulrich}}]{JCD96}
{Christensen-Dalsgaard}, J., {Dappen}, W., {Ajukov}, S.~V., {et~al.} 1996,
  Science, 272, 1286

\bibitem[{{Cunto} {et~al.}(1993){Cunto}, {Mendoza}, {Ochsenbein}, \&
  {Zeippen}}]{Cunto93}
{Cunto}, W., {Mendoza}, C., {Ochsenbein}, F., \& {Zeippen}, C.~J. 1993, \aap,
  275, L5

\bibitem[{{Escobar} {et~al.}(2012){Escobar}, {Th\'eado}, {Vauclair}, {Ballot},
  {Dolez}, {Vauclair}, {Gizon}, {Mathur}, {Quirion}, \& {Stahn}}]{Escobar12}
{Escobar}, M., {Th\'eado}, S., {Vauclair}, S., {et~al.} 2012, \aap submitted

\bibitem[{{Grevesse} \& {Noels}(1993)}]{Grevesse93}
{Grevesse}, N. \& {Noels}, A. 1993, in Origin and Evolution of the Elements,
  ed. {N.~Prantzos, E.~Vangioni-Flam, \& M.~Casse}, 15--25

\bibitem[{{Hui-Bon-Hoa}(2008)}]{Hui08}
{Hui-Bon-Hoa}, A. 2008, \apss, 316, 55

\bibitem[{{LeBlanc} \& {Alecian}(2004)}]{LeBlanc04}
{LeBlanc}, F. \& {Alecian}, G. 2004, \mnras, 352, 1329

\bibitem[{{Mayor} \& {Queloz}(1995)}]{Mayor95}
{Mayor}, M. \& {Queloz}, D. 1995, \nat, 378, 355

\bibitem[{{Michaud}(1970)}]{Michaud70}
{Michaud}, G. 1970, \apj, 160, 641

\bibitem[{{Michaud} {et~al.}(1976){Michaud}, {Charland}, {Vauclair}, \&
  {Vauclair}}]{MichaudMiChVaetal1976}
{Michaud}, G., {Charland}, Y., {Vauclair}, S., \& {Vauclair}, G. 1976, \apj,
  210, 447

\bibitem[{{Michaud} {et~al.}(1978){Michaud}, {Martel}, \& {Ratel}}]{Michaud78}
{Michaud}, G., {Martel}, A., \& {Ratel}, A. 1978, \apj, 226, 483

\bibitem[{{Montmerle} \& {Michaud}(1976)}]{Montmerle76}
{Montmerle}, T. \& {Michaud}, G. 1976, \apjs, 31, 489

\bibitem[{{Paquette} {et~al.}(1986){Paquette}, {Pelletier}, {Fontaine}, \&
  {Michaud}}]{Paquette86}
{Paquette}, C., {Pelletier}, C., {Fontaine}, G., \& {Michaud}, G. 1986, \apjs,
  61, 177

\bibitem[{{Richard} {et~al.}(2001){Richard}, {Michaud}, \&
  {Richer}}]{Richard01}
{Richard}, O., {Michaud}, G., \& {Richer}, J. 2001, \apj, 558, 377

\bibitem[{{Richard} {et~al.}(1996){Richard}, {Vauclair}, {Charbonnel}, \&
  {Dziembowski}}]{Richard96}
{Richard}, O., {Vauclair}, S., {Charbonnel}, C., \& {Dziembowski}, W.~A. 1996,
  \aap, 312, 1000

\bibitem[{{Richer} {et~al.}(2000){Richer}, {Michaud}, \& {Turcotte}}]{Richer00}
{Richer}, J., {Michaud}, G., \& {Turcotte}, S. 2000, \apj, 529, 338

\bibitem[{{Rogers} \& {Nayfonov}(2002)}]{Rogers02}
{Rogers}, F.~J. \& {Nayfonov}, A. 2002, \apj, 576, 1064

\bibitem[{{Schatzman}(1969)}]{SchatzmanSc1969}
{Schatzman}, E. 1969, \aap, 3, 331

\bibitem[{{Schlattl}(2002)}]{Schlattl02}
{Schlattl}, H. 2002, \aap, 395, 85

\bibitem[{{Seaton}(1997)}]{Seaton97}
{Seaton}, M.~J. 1997, \mnras, 289, 700

\bibitem[{{Seaton}(2005)}]{Seaton05}
{Seaton}, M.~J. 2005, \mnras, 362, L1

\bibitem[{{Seaton} {et~al.}(1992){Seaton}, {Zeippen}, {Tully}, {Pradhan},
  {Mendoza}, {Hibbert}, \& {Berrington}}]{Seaton92}
{Seaton}, M.~J., {Zeippen}, C.~J., {Tully}, J.~A., {et~al.} 1992, \rmxaa, 23,
  19

\bibitem[{{Th{\'e}ado} {et~al.}(2009){Th{\'e}ado}, {Vauclair}, {Alecian}, \&
  {LeBlanc}}]{Theado09}
{Th{\'e}ado}, S., {Vauclair}, S., {Alecian}, G., \& {LeBlanc}, F. 2009, ApJ,
  704, 1262

\bibitem[{{Th{\'e}ado} {et~al.}(2005){Th{\'e}ado}, {Vauclair}, {Castro},
  {Charpinet}, \& {Dolez}}]{Theado05}
{Th{\'e}ado}, S., {Vauclair}, S., {Castro}, M., {Charpinet}, S., \& {Dolez}, N.
  2005, \aap, 437, 553

\bibitem[{{Turcotte} {et~al.}(2000){Turcotte}, {Richer}, {Michaud}, \&
  {Christensen-Dalsgaard}}]{Turcotte00}
{Turcotte}, S., {Richer}, J., {Michaud}, G., \& {Christensen-Dalsgaard}, J.
  2000, \aap, 360, 603

\bibitem[{{Turcotte} {et~al.}(1998){Turcotte}, {Richer}, {Michaud}, {Iglesias},
  \& {Rogers}}]{Turcotte98}
{Turcotte}, S., {Richer}, J., {Michaud}, G., {Iglesias}, C.~A., \& {Rogers},
  F.~J. 1998, \apj, 504, 539

\bibitem[{{Wolszczan} \& {Frail}(1992)}]{Wolszczan92}
{Wolszczan}, A. \& {Frail}, D.~A. 1992, \nat, 355, 145

\end{thebibliography}
\end{document}